\documentclass[reprint,prb,aps]{revtex4-1}

\usepackage{graphicx}
\usepackage{dcolumn}
\usepackage{bm}
\usepackage{physics}

\begin{document}

\preprint{APS/123-QED}

\title{Optical spectra in condensed phases: \\how the medium polarizability affects charge-transfer dyes}

\author{ D.K. Andrea Phan Huu}
\author{Cristina Sissa}
\author{Francesca Terenziani}
\author{Anna Painelli}
 \email{anna.painelli@unipr.it}
\affiliation{%
 Department of Chemistry,
Life Science and Environmental Sustainability
}%


\date{\today}

\begin{abstract}
When designing molecular functional materials, the properties of the active specie, the dye, must be  optimized fully accounting for environmental effects.  Here we present an effective model to account for the spectroscopic effects of the medium electronic polarizability on the properties of charge-transfer dyes. Different classes of molecules are considered and the proposed antiadiabatic approach to solvation is contrasted with the adiabatic approach, currently adopted in all quantum chemical approaches to solvation. Transition frequencies and band-shapes are addressed, and the role of the medium polarizability on symmetry-breaking phenomena is also discussed.
\end{abstract}

\maketitle


\section{Introduction}
The effective design of molecular materials for innovative applications requires the concurrent optimization of the active specie, the dye, and its matrix, a highly non trivial task, since the molecular properties depend, in an intrinsically non-linear way, from the properties of the local environment. 
Charge transfer (CT) dyes composed of electron-donor (D) and acceptor (A) 
moieties connected by $\pi$-conjugated bridges 
find applications in solar cells,\cite{solar}
OLED,\cite{oled1,oled2} non-linear optics,\cite{Prasad1991,zyss1994,TerenzianiReview2008,prasad2008} and 
are interesting model
systems for photoinduced CT.\cite{photoinduced,cv07a}
The presence of low-lying excited states
and of delocalized electrons makes these molecules extremely
responsive to the  local environment.\cite{review2007}
Intermolecular interactions in aggregates, supramolecular complexes
and crystals have 
been discussed in different contexts, underlying how mutually
interacting polarizable and/or polar molecules lead to specific  spectroscopic
features that cannot be reconciled with the standard exciton
models.\cite{prb2003,jacs2003,curcumine,dans2017,brunella1,brunella2,spano1,spano2} But
even in comparatively simple systems, where the dye is dissolved in
dilute solutions, in polymeric matrices or glasses, environmental
effects may be quite
impressive, ranging from the solvatochromism of polar
dyes,\cite{liptay,reichardt} to symmetry breaking phenomena \cite{jacs2006,jacs2010,jpcl2010,vanstryland}

\begin{figure}
\centering
	\includegraphics[width=0.9\linewidth]{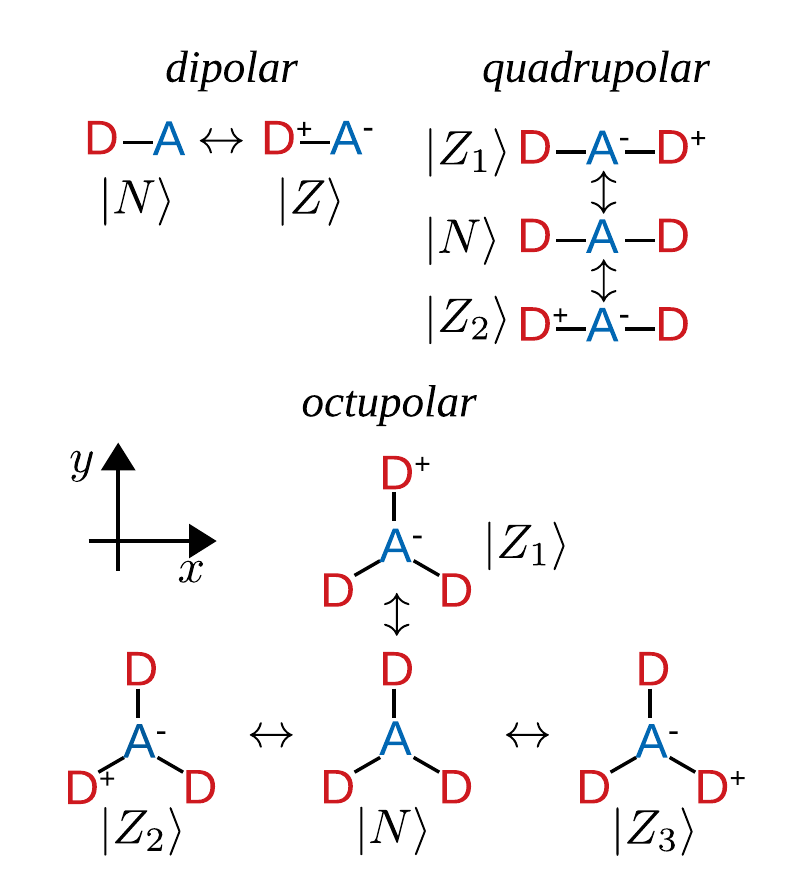}
	\caption{Schematic representation of the main resonating structures of dipolar, quadrupolar and octupolar CT dyes.}
\label{fig:sketch}
\end{figure}

Essential-state models (ESMs) were proposed and successfully applied to
describe low-energy spectral properties of CT dyes in different
environments.\cite{review2007,jacs2006,jacs2010,jpcl2010,dans2017}
ESMs are a family of parametric Hamiltonians that only account for few
electronic molecular states, usually corresponding to the main
resonating structures that characterize each dye, coupled to a few
effective vibrational modes, to account for the geometry relaxation
accompanying the CT process. After the success with dipolar D$-\pi -$A
dyes,\cite{chemphyslett1999,baba} ESMs were applied to more complex quadrupolar and octupolar
structures (see
Fig.~\ref{fig:sketch}).\cite{jacs2006,octupolar,jacs2010} The
main asset of ESMs is the ability to rationalize in a single
theoretical framework linear and non-linear optical spectra of the
dyes, also addressing important and highly non-trivial environmental
effects.

Limiting attention to the simplest environment,
the solvent in a dilute solution, the  most widely discussed
effects are related to the solvent
polarity.\cite{liptay,reichardt,dibella}  A polar solvent stabilizes the
polar  states  of the solute, an effect related to the (re)orientation
of the solvent molecules around the solute. Polar solvation is
therefore related to  slow degrees of freedom.
The best known consequences of polar
solvation are the solvatochromic effects in absorption and
fluorescence spectra of polar dyes,\cite{liptay,reichardt,baba}  as well as
solvent-induced symmetry breaking.\cite{jacs2006,jpcl2010}

The solvation contribution due to  electronic degrees of freedom of the solvent
is much less discussed. It is related to the
electronic polarizability of the solvent, as described by its
refractive index at optical frequencies,  and is then associated to
fast degrees of freedom, that lead to a quasi-instantaneous adjustment of
the electronic cloud of the solvent during an electronic transition of
the solute.  The narrow variability of the refractive index of common
solvents hinders the experimental assessment of this solvation contribution 
and allowed to implicitly account for fast solvation effects in
ESMs via a renormalization of the model parameters, extracted from
experiment.\cite{cp99} However,  fast solvation effects must be properly
understood and explicitly accounted for in order to reliably parametrize ESMs against quantum-chemical
calculations in gas-phase, or to address spectral properties of dyes
in condensed media and, more generally, in media showing
significantly different polarizabilities.

In a recent paper,\cite{prl2020} the role of fast solvation on
spectral properties of organic molecules was addressed underlying how current
approaches to environmental effects, as implemented in
continuum solvation
models\cite{tomasichemrev,chemsci2011,lunkenheimer2012,ciroguido,gabriele2018} 
  as well as
in mixed QM-MM approaches integrating quantum mechanical models for
the molecular species and molecular mechanics models for the
medium,\cite{qmmm} fail to properly address the phenomenon. Here we
extend the discussion to several families of CT dyes, exploiting
ESMs as a simple and effective tool to obtain a
reliable interpretative scheme for solvation effects in these systems. 
Specifically, we will discuss how optical  spectra of CT dyes are affected by the local environment, also addressing the role
of solvation in driving  symmetry breaking in quadrupolar and octupolar dyes.

\section{The model}
\label{section:model}
We consider three different families of dyes, dipolar, quadrupolar and
octupolar systems, whose main resonating structures are depicted in
Fig.~\ref{fig:sketch}. The electronic basis set is defined in all
systems by the neutral structure  $\ket{N}$,
and either 1, 2 or 3 zwitterionic structures  $\ket{Z_i}$ for dipolar,
quadrupolar and octupolar systems, respectively.   We consider
perfectly symmetric quadrupolar and octupolar dyes. Accordingly, for each dye
all zwitterionic structures are equivalent with energy
$2z$ measured with respect to the $|N\rangle$ state, and
are mixed to $\ket{N}$ by  the same matrix element $-\tau$. 
The electronic Hamiltonian reads:
\begin{equation}\label{electronic-hamiltionan}
H_{el}=2z  \hat \rho -\tau\sum_{i=1}^n \ket{Z_i}\bra{N}
\end{equation}
where $i$ runs over the $n$ molecular branches ($n=1$ for dipolar,
$n=2$ for quadrupolar and $n=3$ for octupolar dyes) and 
$\hat{\rho}=\sum_{i=1}^n \ket{Z_i}\bra{Z_i}$ is the ionicity operator
 measuring the cumulative weight 
of zwitterionic structures. 

The dipole moment operator is defined on the diabatic basis, only
accounting for its main contribution, i.e. assigning a sizable dipole of modulus
$\mu_0$ to each zwitterionic $D^+-A^-$ branch. In the reference frame
in Fig.~\ref{fig:sketch}, the dipole moment operator for the different structures
 reads:
\begin{align}
  \label{eq:dipole}
 \hat  \mu_x&=\mu_0\hat{\rho} &\quad & \hat \mu_y=0 &\quad & \hat \mu_z=0  \quad \text{dipolar}\nonumber\\
 \hat \mu_x&=\mu_0\hat{\delta} &\quad & \hat \mu_y=0   &\quad & \hat \mu_z=0  \quad  \text{quadrupolar}\\
    \hat \mu_x&= \mu_0 \hat \delta_x  &\quad & \hat \mu_y=\mu_0 \hat \delta_y  &\quad & \hat \mu_z=0  \quad \text{octupolar} \nonumber
\end{align}
where for the quadrupolar dye we have introduced the auxiliary
operator
\begin{equation}
  \label{eq:delta}
  \hat \delta= \ket{Z_1}\bra{Z_1}-\ket{Z_2}\bra{Z_2}
\end{equation}
and two auxiliary operators are needed for the octupolar dye:
\begin{align}
  \label{eq:deltax}
  \begin{split}
   \hat \delta_x&= \frac{\sqrt{3}}{2} \left(\ket{Z_2}\bra{Z_2}- \ket{Z_3}\bra{Z_3}\right) \\
  \hat \delta_y&= - \ket{Z_1}\bra{Z_1}+\frac{1}{2}\left(
                   \ket{Z_2}\bra{Z_2}+ \ket{Z_3}\bra{Z_3}\right)
    \end{split}
\end{align}

To account for the variation  of the molecular 
geometry upon CT,
  an effective coordinate is introduced, $q_i$ ($p_i$ is the conjugated
momentum) for each molecular arm.
Assuming a harmonic potential with the same curvature for all basis
states  (linear electron-vibration coupling),
the vibrational Hamiltonian reads:
\begin{equation}\label{vibronic-hamiltonian}
H_{mol}=H_{el}-\sqrt{2\epsilon_v}\omega_v\sum_{i=1}^n q_i \ket{Z_i}\bra{Z_i}+\frac{1}{2}\sum_{i=1}^n\left(\omega_v^2 q_i^2+p_i^2\right)
\end{equation}
where $\omega_v$ is the vibrational frequency and $\epsilon_v$ is the
vibrational relaxation energy associated with the CT along each
molecular arm.

For the sake of clarity, when dealing with the interaction between the
dye and the surrounding medium we will use the term solute and solvent
to refer to the dye and the medium, respectively, irrespective of their specific nature. 
 In the simplest model, the solvent  is described as a continuum
 elastic dielectric medium that is perturbed by the solute, described
 as a point dipole. The solvent then generates at the solute location  an
 electric field (the reaction field)  proportional to the solute
dipole. The solute is in turn affected by the
reaction field, leading to a self-consistent problem.\cite{liptay,mcrae,dibella}
The solvent reacts on two different time-scales:  the electronic
solvent  response, with typical frequencies in the UV, is faster than
the solute degrees of freedom. On the opposite,  the orientational
motion of polar solvent molecules  is
much slower than the electronic and vibrational
degrees of freedom of the solute. 
Accordingly, the reaction field $\vec F_R$ is separated into an
 electronic (subscript {\it  el}) and an orientational (subscript {\it or}) contribution,
both proportional to the solute dipole moment:
\begin{equation}
  \label{eq:reaction}
  \vec F_R= \vec F_{el}+ \vec F_{or}=r_{el}\langle \vec \mu\rangle +r_{or}\langle \vec \mu\rangle 
\end{equation}
The $r_{el/or}$ prefactors depend on the dielectric properties of the
medium. Explicit expressions have been derived relating these quantities to the
medium refractive index, dielectric constant and the shape and size of
the cavity occupied by the solute.\cite{mcrae,dibella}
Irrespective of the specific model,
the orientational component of the solvation field is sizable only for
polar solvents, so that $r_{or}\sim 0$ in non-polar media. 

The different timescales of the two components of the reaction
field call for different approximation schemes. Specifically,
the adiabatic approximation
can safely be applied to slow solvation, while fast solvation can be
dealt with in the antiadiabatic approximation,\cite{ciuchi} as 
discussed in Ref. \citenum{prl2020}. Accordingly, the complete Hamiltonian, also accounting for the  solute-solvent interaction, reads
\begin{equation}\label{total-hamiltonian}
H=H_{mol}+\left[ \frac{F_{el}^2}{2r_{el}}+T_{el}-\hat{\vec{\mu}}\cdot
  \vec F_{el} \right]+\left[
  \frac{F_{or}^2}{2r_{or}}-\hat{\vec{\mu}}\cdot \vec F_{or} \right]
\end{equation}
where the terms relevant to electronic and
orientational solvation have been separated into square brackets. In both terms the quadratic contribution in
the reaction field accounts for the potential energy (the energy
required to create the field) in the harmonic approximation. The relevant 
force constant is fixed to $(r_{el/or})^{-1}$ upon imposing the equilibrium
condition in Eq. \ref{eq:reaction}.\cite{cp99,prl2020} The kinetic energy 
relevant to polar solvation is  neglected in the adiabatic
approximation, while it is accounted for as  $T_{el}$ for the
electronic component. 
Polar solvation  has been extensively discussed in the framework of
ESMs and will not be addressed here. 
We therefore only discuss the electronic (fast) contribution to solvation
dropping the last term in the Hamiltonian
in Eq. \ref{total-hamiltonian}, setting $F_{or}=0$, as relevant to 
non-polar solvents.

The effective frequency $\omega_{el}$
associated to the fast component of the reaction field 
accounts for the electronic excitation of the solvent, typically in the mid-far UV regions, at much higher frequencies than the transition frequencies of the solute,  in the visible or near-UV region. 
Accordingly, we adopt an antiadiabatic approximation, setting  $\omega_{el} \rightarrow \infty$,\cite{prl2020}
to define the following renormalized solute Hamiltonian, implicitly accounting for fast solvation: \cite{prl2020}
\begin{equation}
  \label{eq:aa1}
  H_{AA}=H_{mol}-\frac{r_{el}}{2}\hat{\vec \mu}^2
\end{equation}
Fast solvation then introduces a two-electron term in the electronic
Hamiltonian, that however, in ESM acquires a  very simple form with a
clear   physical meaning. Indeed for all CT dyes discussed here,  with dipolar, quadrupolar or
octupolar structure, one finds $\hat{\vec \mu}^2=\mu_0^2 \hat \rho  $,
so that 
\begin{equation}
  \label{eq:aa2}
  H_{AA}=H_{mol}-\epsilon_{el} \hat \rho
\end{equation}
where $\epsilon_{el}=\mu_0^2 r_{el}/2$ measures the amount of energy gained by
the system in a zwitterionic state due the relaxation of
the electronic clouds of the solvent molecule.
In other terms, when going from gas phase
to solution, the energy $2z$ required to separate a charge along a
molecular arm is reduced by the solvent relaxation energy related to
fast solvation, so that  $z \rightarrow z-\epsilon_{el}/2$.

The above equations are fairly general and do not depend on the
details of solvation model. Relating $\epsilon_{el}$ to the
solvent properties and specifically to the solvent refractive index,
requires however a specific description of the solute-solvent
system. A widely adopted approach assumes that the solute occupies a
spherical cavity in the solvent, with radius $a$.\cite{mcrae,dibella} In this hypothesis
\begin{equation}
  \label{eq:erre}
  r_{el}= \frac{2}{4\pi\epsilon_0a^3} \frac{\eta^2-1}{2\eta^2+1}
\end{equation}
where $\eta$ is the solvent refractive index at optical
frequencies. We will use this approximate relation to estimate  a reasonable
variability range for $\epsilon_{el}$.

In the following, we will present selected results obtained for the
three  families of dyes in Fig. \ref{fig:sketch}, to discuss the effects of
the medium polarizability on molecular properties. Relevant results
will be compared  with results  obtained adopting the adiabatic
approximation to describe fast solvation. 
In this approximation,  the kinetic energy associated with the electronic degrees of
freedom of the solvent, $T_{el}$ in Eq. \ref{total-hamiltonian}, is
neglected and an effective  Hamiltonian is defined for each
electronic state of the solute, fixing  $\vec F_{el}$
to its equilibrium value relevant the specific state of interest.
Due to the fast dynamics involved in the solvent electronic
polarization,  the adiabatic approximation is clearly not suitable to
describe fast solvation.\cite{prl2020}
However, current implementations of continuum solvation
models in quantum chemical codes all rely in the adiabatic
approximation for both the orientational and electronic components of
solvation field,\cite{cammi:lr-ss,improta,caprasecca}
and the same approximation is also adopted in current
application of QM-MM models.\cite{qmmm} It is therefore important to stress the
limits of this widely adopted, even if not explicitly acknowledged, approximation.

\section{RESULTS AND DISCUSSION}

\subsection{Polar dyes}
\begin{figure*}
\centering
\includegraphics[width=0.49\textwidth]{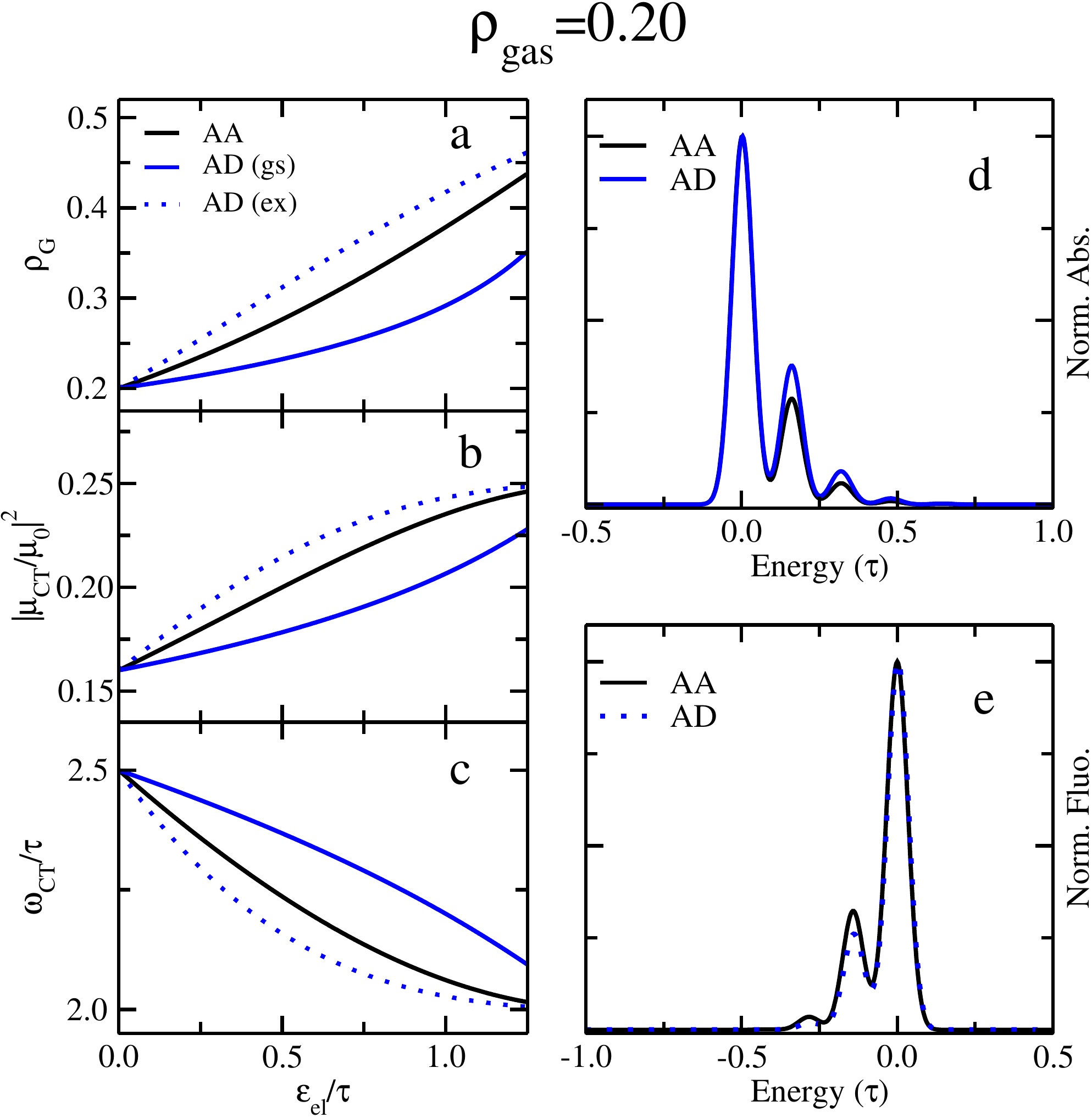}
\includegraphics[width=0.49\textwidth]{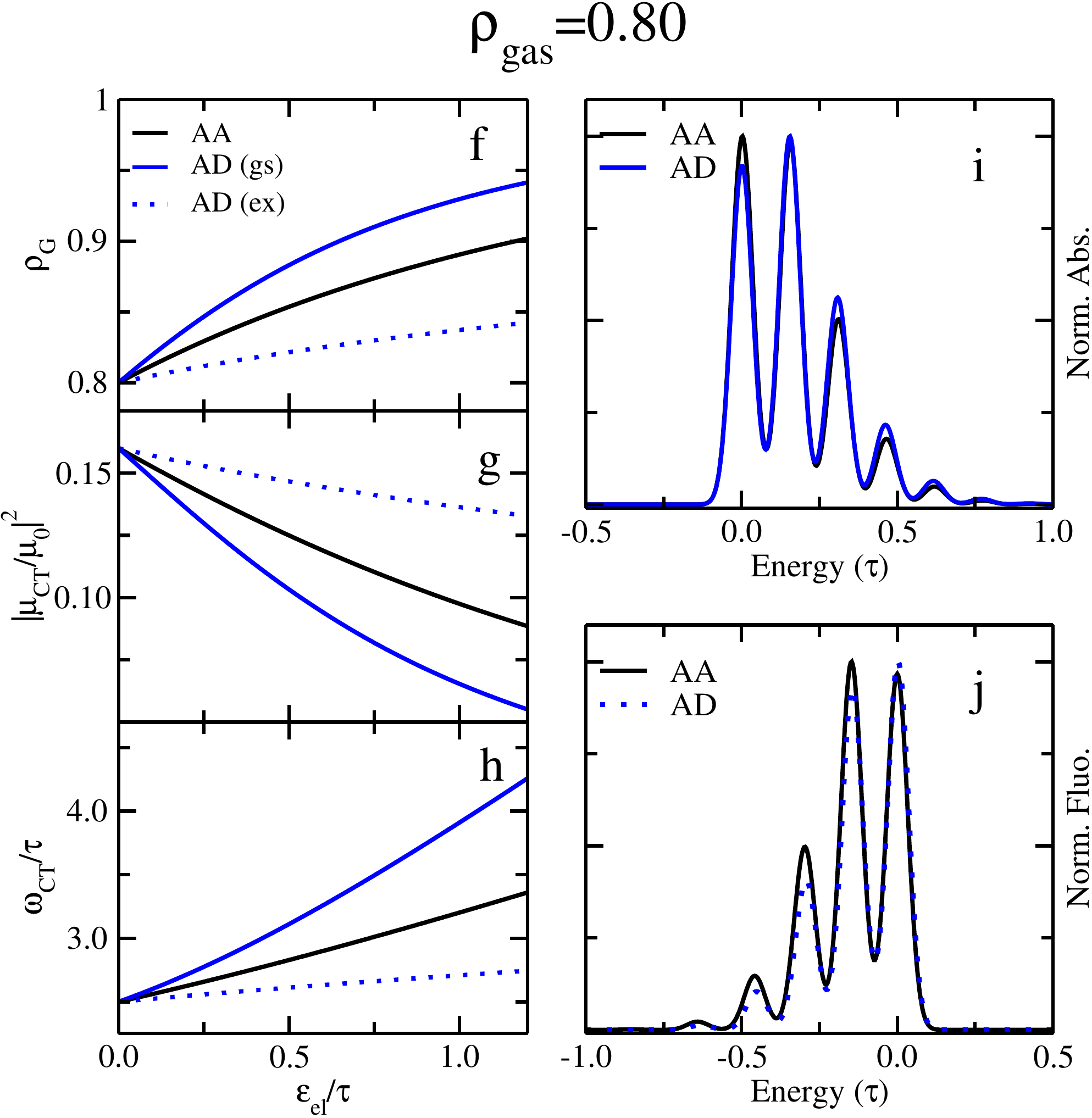}\\
\caption{Fast solvation effects on the properties  of polar
  chromophores with $\rho_{gas}=0.2$ (left side) or  
$\rho_{gas}=0.8$ (right side). On each side, right panels show 
properties (ground-state ionicity, transition dipole moment,
transition frequency)  as a function of $\epsilon_{el}$ for a system with $\epsilon_{v}=0$ . 
The right panels show vibronic bandshapes calculated for a system with
$\epsilon_{v}=0.3$ and $\epsilon_{el}=0.32$.
Antiadiabatic (AA) results: black lines; Adiabatic (AD) results with $F_{el}$ equilibrated with the ground state (full blue lines) or with the optically-allowed excited state (dotted blue lines).}
\label{fig:dipolar}
\end{figure*}
Neglecting electron-vibration coupling ($\epsilon_{v}=0$ in
Eq. \ref{vibronic-hamiltonian}), the molecular properties of  polar D$-\pi -$A dyes only depend on the $z/\tau$
ratio. In the following we use units such that  $\hbar=1$ and set $\tau$ as the energy unit. 
The  actual $\tau$ value for most CT dyes is of the order of 1 eV,\cite{baba}
even if for  dyes of interest for thermally delayed fluorescence 
applications,\cite{oled2,adachi,monkman} typical $\tau$ are
one order of magnitude smaller.
All properties of interest can be expressed as a function of $\rho$:
the transition dipole moment is  $\mu_{CT}=\mu_0 \sqrt{\rho(1-\rho)}$,
showing  a maximum  at $\rho=0.5$ , and the transition frequency is
$\omega_{CT} =\tau / \sqrt{\rho(1-\rho)}$,  showing  a minimum  at $\rho=0.5$.
The permanent dipole moment is $\mu_G=\mu_0 \rho$ and  $\mu_E=\mu_0 (1-\rho)$ in
the ground and excited state, respectively, so that the mesomeric
dipole moment $\mu_E-\mu_G=\mu_0(1-2\rho)$ is positive for mostly
neutral dyes ($\rho<0.5$) and negative for mostly zwitterionic dyes
($\rho>0.5$).  Accordingly,
mostly neutral dyes show  normal solvatochromic
behavior (the absorption band redshifts upon  increasing solvent
polarity) while mostly zwitterionic dyes show inverse solvatochromism
(the absorption band blueshifts upon increasing solvent polarity), so
that a simple spectroscopic data allow to discriminate between the
two classes of polar dyes.

Fig.~\ref{fig:dipolar} collects results for a mostly neutral dye
($z=0.75$, corresponding to $\rho_{gas}=0.2$) and a mostly 
zwitterionic dye ($z =-0.75$, $\rho_{gas}=0.8$).
Analytical results for the purely electronic model ($\epsilon_v=0$)
in panels a-c and f-h show the evolution of the molecular properties
when $\epsilon_{el}$ increases from 0, as relevant to the gas phase,
to larger values, typical of organic media. In all cases,  the
ionicity $\rho$ increases with $\epsilon_{el}$, as a result of the
stabilization of charge separated states by the electronic
polarizability of the  environment. Adiabatic results (blue lines in Fig. \ref{fig:dipolar}) show
quantitative and sometimes sizable deviations from the antiadiabatic
results (black curves), but the most clear failure of the adiabatic approach is
recognized in having two different sets of results, corresponding to
the two different adiabatic Hamiltonians obtained upon fixing the
reaction field to the equilibrium value for the ground state (blue continuous lines) or for the
excited state (blue dotted lines). This is clearly unphysical, since the electronic clouds
of solvent molecules readjust quickly (instantaneously in the
antiadiabatic limit) to the charge reorganization in the solute.
In any case, taking adiabatic results at 
face value, one should use the adiabatic Hamiltonian with $F_{el}$
equilibrated to the ground state or to the excited state to simulate
absorption or fluorescence processes, respectively. This leads
to a spurious red-shift of the fluorescence band with respect
to the absorption band. To get rid of this spurious Stokes-shift,
two different approaches are currently implemented in quantum chemical
packages. In the first approach, first order perturbation theory 
is used to calculate the energy corrections due to solvation, without
the need to diagonalize two different
Hamiltonians.\cite{chemsci2011,ciroguido}
In the second
approach 
the transition energy (coincident for absorption and fluorescence)
is defined as the difference between the energy of the excited state, calculated
diagonalizing  the adiabatic Hamiltonian with $F_{el}$ equilibrated to the
excited state, and the energy of the ground state, calculated diagonalizing
the adiabatic Hamiltonian with $F_{el}$ equilibrated to the  ground-state.\cite{improta2006} 
This strategy is however untenable on physical grounds, as it defines
electronic transitions between states obtained upon diagonalizing
two different Hamiltonians. The inconsistency of the approach, leading
to undefined transition dipole moments, most clearly points to the
fundamental failure of the adiabatic approach to fast
solvation. 

To address vibronic bandshapes  we account for electron-vibration coupling in a non-adiabatic calculation. Specifically, we solve the molecular Hamiltonian in Eq. 5 (modified as in Eq. 8 to
account for fast solvation in the antiadiabatic limit, or accounting
for the static corrections due to the equilibrated $F_{el}$ in the
adiabatic limit) by writing the corresponding matrix on the
basis defined as the direct product of the electronic basis states
times the eigenstates of the harmonic oscillator(s) in the last term
of equation 5. Of course the vibrational basis is truncated to a large
enough number of vibrational states as to obtain convergence. Once the molecular
Hamiltonian is diagonalized, the absorption and fluorescence spectra
are  calculated from the  transition energies and transition
dipole moments assigning  each vibronic transition a  Gaussian
lineshape with fixed linewidth (in this work it is set to  0.04).\cite{annine} 
In the antiadiabatic approach to fast solvation the eigenstates obtained upon diagonalization of 
a single effective Hamiltonian enter the calculation of  both absorption and fluorescence
spectra. On the opposite, in the adiabatic
approach to fast solvation two different Hamiltonians with $F_{el}$ equilibrated either to the ground or to
the excited state are used for the calculation of absorption and
fluorescence spectra, respectively.
In Fig.~\ref{fig:dipolar}, the rightmost panels relevant to each dye show an example of vibronic bandshapes
calculated for absorption and emission spectra setting $\epsilon_v=0.3$. 
 Since we are interested in comparing bandshapes, all normalized
 spectra  are translated to set the origin of the
 energy axis at the  0-0 transition energy. Calculated absorption and
 fluorescence bandshapes in
 the adiabatic approximation are marginally different from
 antiadiabatic results.
 
\subsection{Quadrupolar chromophores}
\begin{figure}
\centering
\includegraphics[width=1.0\linewidth]{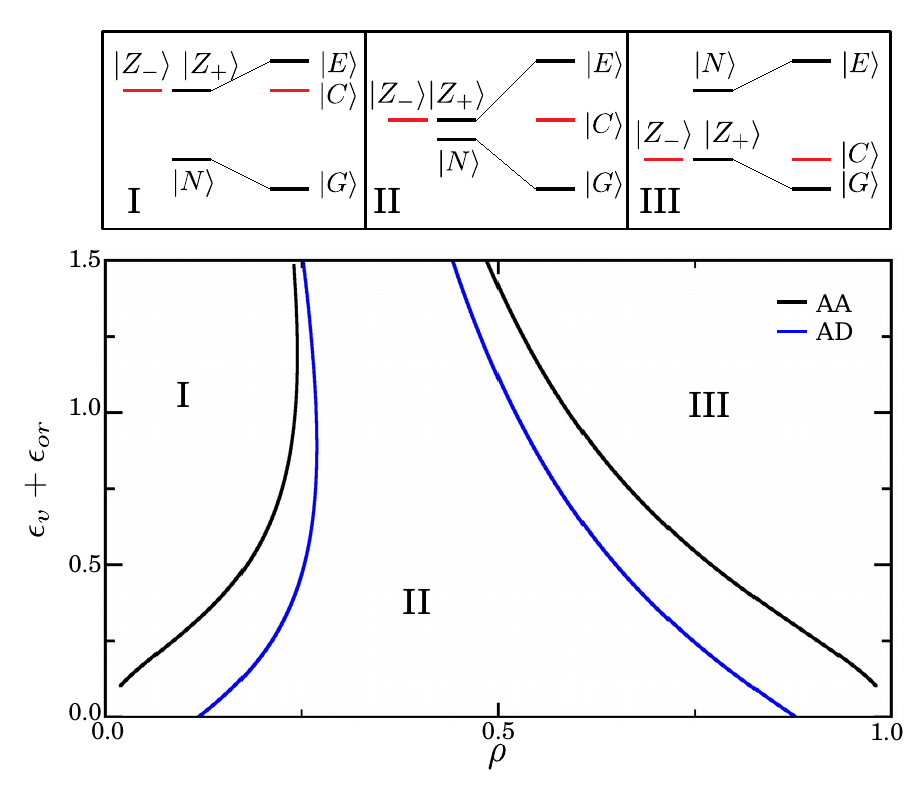}
\caption{Top panels: sketch of the essential states of class I, II and
  III quadrupolar dyes. Bottom panel:  phase diagram for quadrupolar
  dyes. The black line show antiadiabatic (AA) results, the blue lines
  show   adiabatic (AD) results for $\epsilon_{el}=0.3$.}
\label{fig:phase-diagram}
\end{figure}
To address the electronic problem of quadrupolar dyes we exploit
symmetry, combining the
two degenerate basis states $|Z_1\rangle$ and $|Z_2\rangle$
as $|Z_\pm\rangle=( |Z_1\rangle \pm |Z_2\rangle)/\sqrt{2}$). The mixing between 
$|Z_+\rangle$ and  $|N\rangle$ gives the ground state
$|G\rangle$ and an excited  state $|E\rangle$. The excited state
$|E\rangle$ cannot be reached upon one-photon
absorption and is  located at higher energy than the
optically active state $|C\rangle=|Z_-\rangle$. The mixing between $|N\rangle$
and $|Z_+\rangle$ only depend on the $z/\tau$ ratio.\cite{jacs2006} 
As sketched in the top panels of
Fig.~\ref{fig:phase-diagram}, systems with large mixing ($\rho
\sim 0.5$) are characterized by large transition energies   (class II
dyes),
while systems with small mixing ($\rho
\rightarrow$ 0 or 1 class I or III dyes, respectively) show a pair of
quasi-degenerate eigenstates, signalling a conditional instability.\cite{jacs2006}

Studying the problem for the isolated molecule in the gas phase
(Eq. 5) we can collect valuable information about the tendency of the
dye towards symmetry breaking, adopting an adiabatic approximation to
treat molecular vibrations. Along these lines, the potential energy
surfaces (PES) for the ground and excited states can be drawn and analytical
results may be obtained for the phase diagram of quadrupolar dyes
(bottom panel of Fig.~\ref{fig:phase-diagram}). In the $(\epsilon_v,
\rho)$ plane the black curves  mark the boundaries between the three
different classes: for class I dyes the PES associated to the first
excited state  shows a
double minimum, suggesting the tendency to symmetry breaking for this state.
Class II dyes are characterized by well-behaved PES for all three
states that are therefore not prone to symmetry-breaking. Finally
class III dyes are characterized by a bistable ground state. It is
important to underline at this stage that symmetry breaking cannot be
observed in an isolated molecule.\cite{anderson} The double minimum in
the excited or ground state of systems in class I or III, respectively,
does not necessarily imply a symmetry breaking phenomenon, since any finite-size
system will always oscillate between the two minima recovering the
full symmetry of the system in a sort of dynamical Jahn-Teller
effect.\cite{dimaiolo-jcp} Of course a genuine symmetry breaking may
be observed if the dye is dissolved in a polar solvent. Polar
solvation, corresponding to an extremely slow motion, can be described
accurately in the adiabatic approximation, and the relevant relaxation
energy, $\epsilon_{or}= \mu_0^2r_{el}/2$, enters the picture summing
up to $\epsilon_v$ in the phase diagram in
Fig. ~\ref{fig:phase-diagram}, thus widening the region where either the
ground or excited state instability occurs.\cite{jacs2006} Even more importantly, the
slow motion of polar solvation basically freezes the system in one of
the minima not allowing tunneling in any time of relevance to optical
spectroscopy. Symmetry breaking driven by polar solvation in the
excited state of class I polar dyes quite naturally explains the large
positive solvatochromism observed in
fluorescence spectra of these  systems,\cite{jacs2006,vanstryland}
while the ground-state symmetry breaking  in class III dyes is the key
to understand the anomalous absorption solvatochromism observed in long cyanine dyes, in
spite of their nominally symmetric structure.\cite{jpcl2010,sissa:cyanine}
\begin{figure*}
\centering
\includegraphics[width =0.75\textwidth]{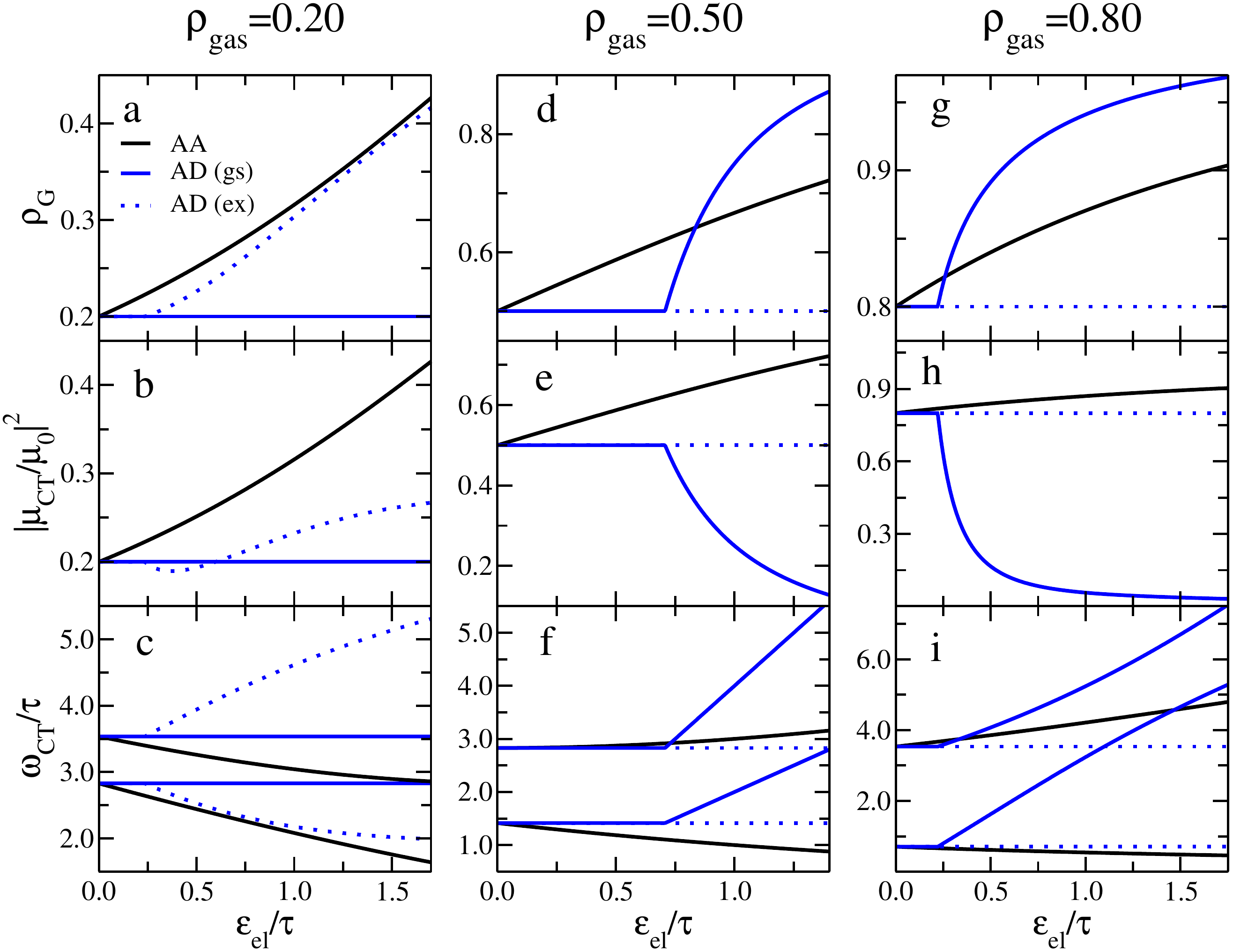}
\caption{Fast solvation effects on the properties (ground-state
  ionicity, transition dipole moment, transition frequency) of
  quadrupolar dyes belonging to different classes and with $\epsilon_v=0$. Two transition
  energies are shown corresponding to the C and E states (lower and
  higher transition energy, respectively). The transition dipole
  relevant to the transition from the ground to the E state vanishes
  and is not shown. Antiadiabatic (AA) results: black lines; Adiabatic
  (AD) results with $F_{el}$ equilibrated with the ground state (full
  blue lines) or with the optically-allowed excited state (dotted blue
  lines). The abrupt changes observed in adiabatic results mark the
  occurence of symmetry breaking.}
\label{fig:quad-electronic}
\end{figure*}

Accounting for fast solvation does not alter the picture. 
Indeed in the antiadiabatic approximation, the electronic
polarizability of the solvent lowers the  energy gap $2z$,
as discussed in Section~\ref{section:model}, leading to an 
increase of $\rho$. However, the phase diagram in
Fig.~\ref{fig:phase-diagram} still applies:  the black lines separating the
different regions in the phase diagram are not affected by the
variation of the medium refractive index. Instead,  if  the
adiabatic approximation is incorrectly enforced to describe fast
solvation,  the relevant relaxation energy
$\epsilon_{el}$ would enter the picture much as in the case of polar
solvation, hence summing up to  $\epsilon_{v}$ in
Fig. \ref{fig:phase-diagram}  favoring
symmetry breaking. In other terms, as illustrated  in
the phase diagram in Fig.~\ref{fig:phase-diagram} for the specific
case $\epsilon_{el}=0.3$ (blue lines), the boundaries between the
different regions in the phase diagram would be downshifted by
$\epsilon_{el}$, artificially widening the instability regions
associated with class I and class III dyes. 
 \begin{figure*}
\centering
\includegraphics[width =0.7\textwidth]{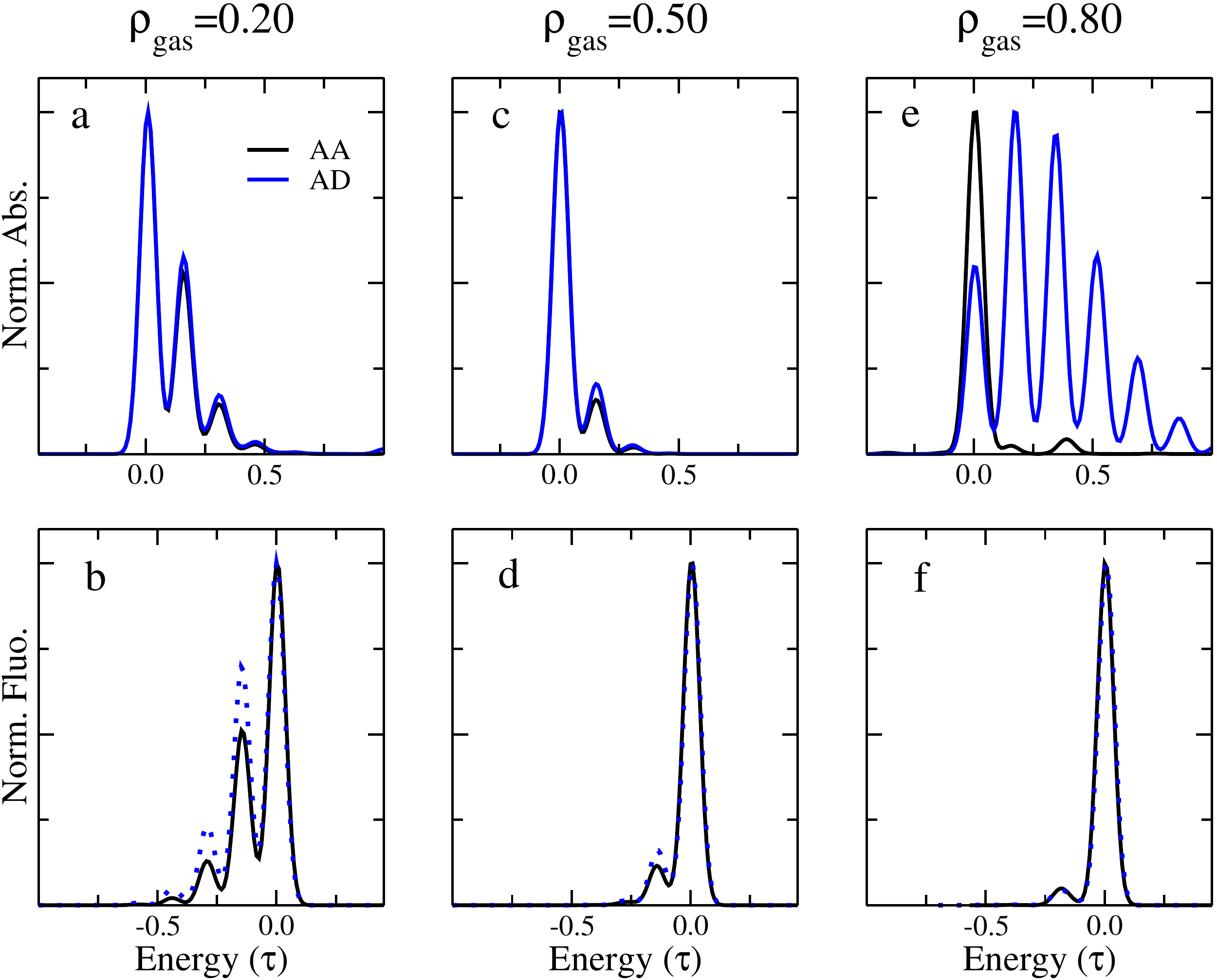}
\caption{Vibronic absorption (top) and fluorescence (bottom) spectra
  for quadrupolar dyes  of different classes, for $\epsilon_{v}=0.3$
  and $\epsilon_{el}=0.32\tau$. Black lines: antiadiabatic results;
  Blue lines: adiabatic results.}
\label{fig:quad-bandshapes}
\end{figure*}

Fig.~\ref{fig:quad-electronic} shows analytical results for the
electronic model ($\epsilon_v=\epsilon_{or}=0$) relevant to three quadrupolar systems
with $z$ adjusted as to have $\rho_{gas}=$~0.2, 0.5 and 0.8,
as representative of class I, II and III dyes, respectively. In all cases, the
antiadiabatic results (black lines) predict an increase of $\rho$ with
increasing the medium polarizability. This always implies an increase
of the transition dipole moment $\mu_{CT}$ for the allowed
$G\rightarrow C$ transition. For the class I system ($\rho_{gas}=0.2$) the
two transition frequencies ($G\rightarrow C$ and $G\rightarrow E$) decrease    
considerably with the medium refractive index, while the effects are
less pronounced in the other two systems, with the lowest (highest) transition
decreasing (increasing) in energy with $\epsilon_{el}$. 

Enforcing the adiabatic approximation for fast solvation
leads to different Hamiltonians, depending on the reference state
selected to equilibrate the reaction field. Continuous and dotted blue lines in
Fig.~\ref{fig:quad-electronic} refer to adiabatic results obtained fixing $F_{el}$ to the
equilibrium value relevant to the ground state or to the optically-allowed ($C$)
excited state, respectively. Since the ground-state dipole moment vanishes as long
as the ground state symmetry is conserved, adiabatic results obtained
for $F_{el}$ equilibrated to the ground state do not vary at all with
$\epsilon_{el}$ as long as the ground-state symmetry is preserved. This is the case for the quadrupolar dye with  $\rho_{gas}=0.2$ (left panels), where no variation of either $\rho_G$ or $\mu_{CT}$
or $\omega_{CT}$  is obtained in the ground-state adiabatic approximation (full blue lines) when  $\epsilon_{el}$ is increased.
On the other hand,  adiabatic results obtained
for $F_{el}$ equilibrated to the optically-allowed excited state do not vary at all with
$\epsilon_{el}$ as long as the excited-state symmetry is preserved. This is the case of
 the quadrupolar dye with  $\rho_{gas}=0.8$ (right panels), where no variation of either $\rho_G$ or $\mu_{CT}$
or $\omega_{CT}$  is obtained in the excited-state adiabatic
approximation (dotted blue lines) when  $\epsilon_{el}$ is
increased. The adiabatic results contrast sharply with antiadiabatic
results that instead properly account for the effect of the solvent
polarizability on molecular properties. 

However, the most striking failure of the  adiabatic approximation to
fast solvation in quadrupolar systems is the prediction of  spurious
symmetry-breaking phenomena. The class II system in the middle panels
of  Fig.~\ref{fig:phase-diagram} ($\rho_{gas}=0.5$) is a paradigmatic
example:  if electronic solvation is properly
described in the antiadiabatic approximation, the system is not prone
to symmetry breaking, but, if a ground state adiabatic
approach is enforced, the ground state undergoes symmetry breaking at
$\epsilon_{el}\approx 0.75$, as shown by the abrupt variation of the molecular
properties (full blue lines). At the same time, symmetry is preserved in the excited
state, so that in the adiabatic approximation, when $F_{el}$ is
equilibrated to the $C$ state,  $\epsilon_{el}$ does not affect
molecular properties  (dotted blue lines). 
For the system with $\rho_{gas}=0.8$, the ground-state adiabatic
approximation predicts symmetry breaking in the ground state for
$\epsilon_{el} > 0.25$, while symmetry is preserved in the $C$ state.  
For the system with $\rho_{gas}=0.2$, the excited-state adiabatic
approximation predicts symmetry breaking in the $C$ state for
$\epsilon_{el} > 0.30$, while symmetry is preserved in the ground state.

Vibronic bandshapes are shown in Fig.~\ref{fig:quad-bandshapes} for the same three representative systems, but fixing 
$\epsilon_v=0.3$ and $\epsilon_{el} = 0.32$. Marginal differences
between the antiadiabatic and adiabatic results are found as long as
symmetry is conserved in the adiabatic calculation, while sizable
deviations are of course  observed for emission spectra of class I
dyes and huge deviations for absorption spectra of class III dyes, due
to spurious symmetry-breaking effects. 
\begin{figure}
\centering
\includegraphics[width=1.0\linewidth]{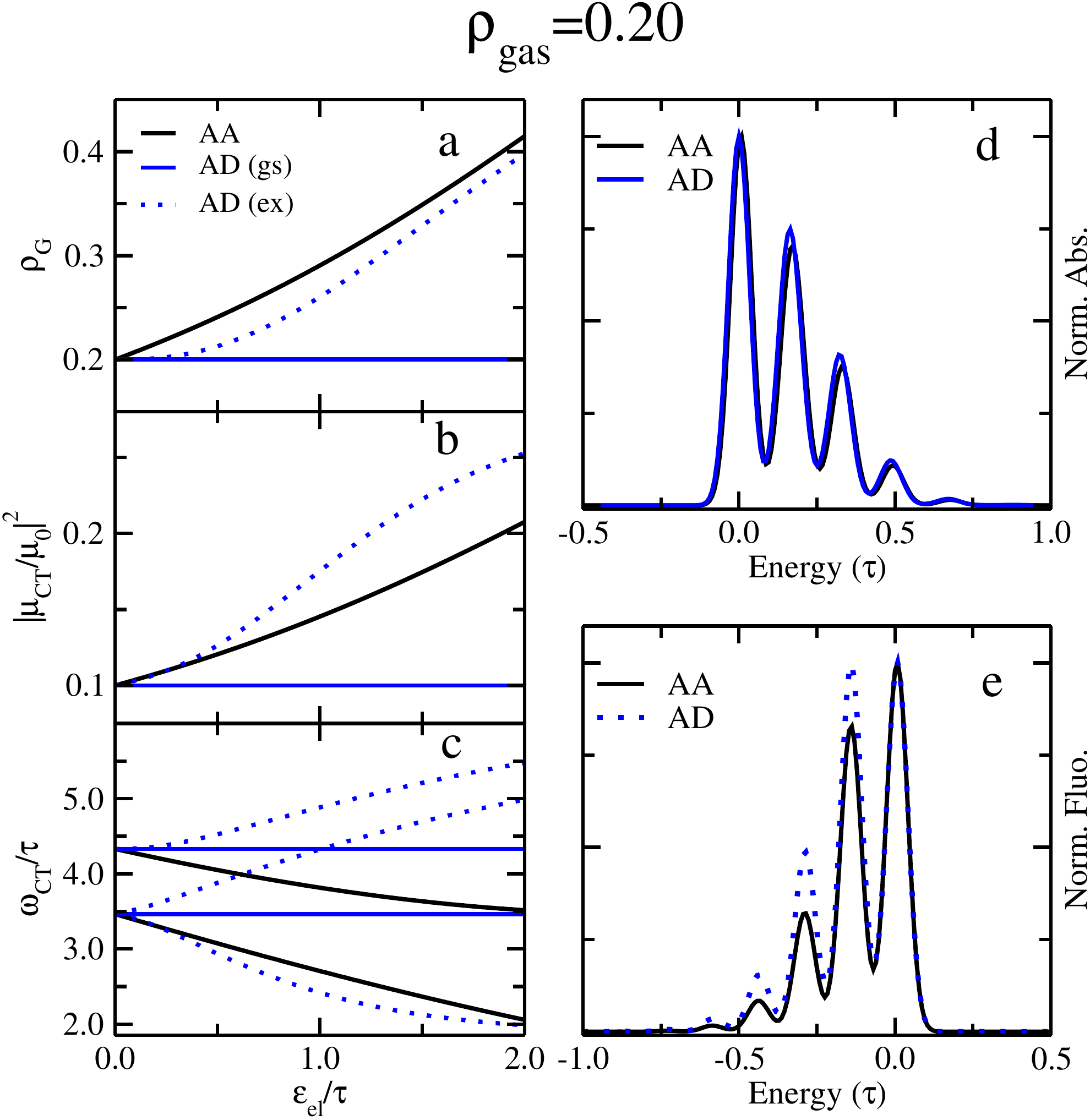}
\includegraphics[width=.75\linewidth, trim =0 0 0 0, clip]{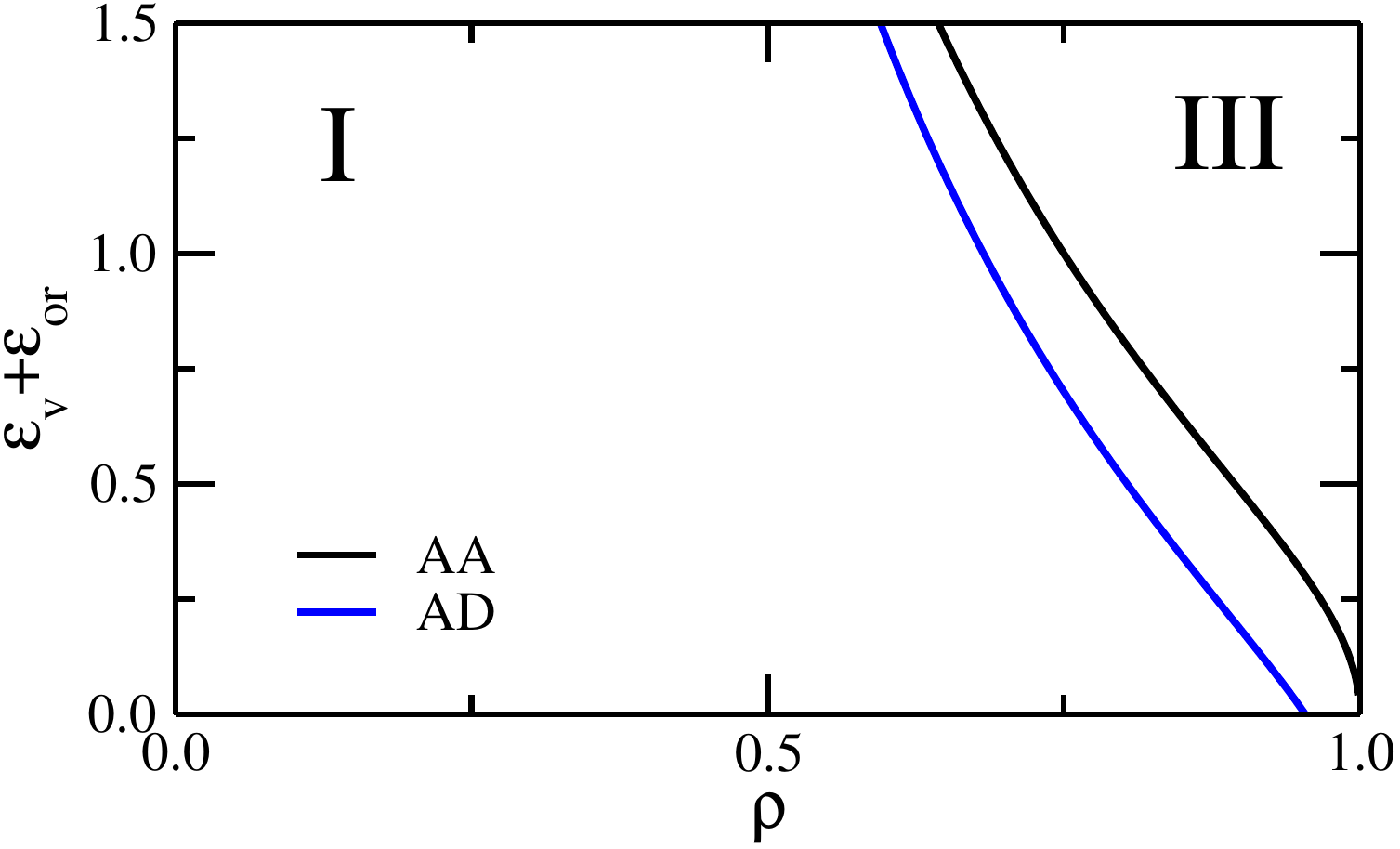}
\caption{Top panels: fast solvation effects on an octupolar chromophore with
  $\rho_{gas}=0.20$.  Black lines show antiadiabatic (AA) results;
  blue lines show adiabatic (AD) results with $F_{el}$ equilibrated
  with the ground state (continuous lines) or with the
  optically-allowed excited state (dotted  lines). Panels (a-c) show the electronic properties (ground-state ionicity, transition dipole
  moment, transition frequencies) as a function of
  $\epsilon_{el}$ for a system with $\epsilon_v=0$. 
  Panels (d) and (e) show vibronic absorption and fluorescence
  spectra, respectively,  calculated for $\epsilon_{el}=0.32$ and $\epsilon_v=0.3$.
  Bottom panel: phase diagram ($\epsilon_{el}=0.3$ for the adiabatic, AD, result).}
\label{fig:octupolar}
\end{figure}



\subsection{Octupolar chromophores}

The threefold rotation axis in octupolar chromophores implies the
presence of doubly degenerate states, which leads to instability in
either the ground or excited state, precluding the presence of  class
II dyes, as shown in the phase diagram in the bottom panel of  Fig.~\ref{fig:octupolar}.\cite{octupolar} As in the case of quadrupolar chromophores, the phase diagram, plotted against the ground-state ionicity, is independent of $\epsilon_{el}$ in the correct antiadiabatic limit (black line). In the adiabatic approximation instead the boundary is lowered along the ordinate by $\epsilon_{el}$ (blue line).

The left panels of Fig.~\ref{fig:octupolar} show the molecular properties calculated for an octupolar dye with $\rho_{gas}=0.2$ 
(to the best of our knowledge there are no examples of octupolar dyes of class III). 
The antiadiabatic calculation predicts, as expected, an increasing contribution of zwitterionic states into the ground state (increasing
$\rho$) when $\epsilon_{el}$ is increased. Concomitantly, the transition dipole moment towards
the optically-allowed state, corresponding to a doubly degenerate state,
increases while the excitation energy towards either  the lowest-energy
(allowed) or the highest-energy (forbidden) excited states decreases.
The system stays stable, preserving its symmetry, as long as slow degrees of freedom do not enter into play. The adiabatic calculation instead
predicts no effect of the medium polarizability when $F_{el}$ is
equilibrated to the ground state. On the opposite, when the reaction
field is equilibrated to the lowest excited state, clear signatures of
a spurious symmetry breaking appear.  
As for vibronic bandshapes, results in Fig. \ref{fig:octupolar}(d) and
(e) for the dye with $\rho_{gas}=0.2$, $\epsilon_v=0.3$ and
$\epsilon_{el}=0.32$
show marginal
differences between spectra calculated in the adiabatic vs the antiadiabatic approximation.

\section{CONCLUSIONS}
Solvation is a complex phenomenon involving  several degrees of
freedom characterized by different timescales. In particular, a slow
component of solvation is driven by the orientational motion of polar 
solvent molecules around the solute, and is only relevant to polar
solvents.  Another component is instead always present,
irrespective of the solvent polarity, and is related to the solvent
electronic polarizability, as measured by the solvent refractive
index. This corresponds to a fast motion, since the electronic
excitations of the  solvent typically fall in the mid/far-UV region,
i.e. at significantly higher frequencies than relevant  degrees
of freedom of organic dyes.  
While the slow component of solvation can be safely dealt with  in the
adiabatic approximation, the
same approximation is not suitable to treat fast  solvation, that can instead 
be treated in the antiadiabatic approximation.\cite{prl2020}
Here we discussed spectroscopic effects of  fast solvation with
reference to ESMs for CT dyes of different
families. In particular, we discussed how  the
medium polarizability affects optical spectra of  dipolar, quadrupolar and octupolar dyes,
comparing  results obtained in the antiadiabatic approximation with
those  obtained in an adiabatic approach.  
This is not a trivial exercise since   current implementations of solvation
models and more generally all quantum-classical description of
molecules in solution rely on an  adiabatic treatment of fast
solvation. Apart from quantitative deviations, the adiabatic
approximation to fast solvation leads to  spurious Stokes shifts
(measuring the difference between the energy of the absorption and
emission transitions) that are currently cured either
limiting the treatment to a first order perturbation approach or
accounting for a
transition occurring between states obtained as eigenstates of two
different Hamiltonians.

Moreover, in quadrupolar and octupolar dyes, applying the 
adiabatic approximation to fast solvation can drive symmetry
breaking in systems where it cannot possibly occur. Specifically,
genuine symmetry breaking can never occur in isolated (gas phase)
molecular systems,\cite{anderson} nor can it be induced by fast
solvation. Only polar solvation, associated with an extremely slow,
classical coordinate may drive a \textit{bona fide} symmetry breaking in a
molecular system. Symmetry breaking phenomena as often discussed in chemical
literature\cite{terenziani2005a,terenziani2005b,jacquemin2020,dongho2020} in the gas-phase or in non-polar solvents 
are actually an artifact associated with the adiabatic treatment of
vibrational degrees of freedom and/or of fast solvation.

\begin{acknowledgments}
This project received funding from the European Union Horizon 2020 research and innova-
tion programme under Grant Agreement No. 812872 (TADFlife), and benefited from the
equipment and support of the COMP-HUB Initiative, funded by the ‘Departments of Excellence’ program of the Italian Ministry for Education, University and Research (MIUR,
2018-2022). We acknowlegde the support from the HPC (High Performance Computing)
facility of the University of Parma, Italy.
\end{acknowledgments}

\bibliography{biblio}

\begin{thebibliography}{52}%
\makeatletter
\providecommand \@ifxundefined [1]{%
 \@ifx{#1\undefined}
}%
\providecommand \@ifnum [1]{%
 \ifnum #1\expandafter \@firstoftwo
 \else \expandafter \@secondoftwo
 \fi
}%
\providecommand \@ifx [1]{%
 \ifx #1\expandafter \@firstoftwo
 \else \expandafter \@secondoftwo
 \fi
}%
\providecommand \natexlab [1]{#1}%
\providecommand \enquote  [1]{``#1''}%
\providecommand \bibnamefont  [1]{#1}%
\providecommand \bibfnamefont [1]{#1}%
\providecommand \citenamefont [1]{#1}%
\providecommand \href@noop [0]{\@secondoftwo}%
\providecommand \href [0]{\begingroup \@sanitize@url \@href}%
\providecommand \@href[1]{\@@startlink{#1}\@@href}%
\providecommand \@@href[1]{\endgroup#1\@@endlink}%
\providecommand \@sanitize@url [0]{\catcode `\\12\catcode `\$12\catcode
  `\&12\catcode `\#12\catcode `\^12\catcode `\_12\catcode `\%12\relax}%
\providecommand \@@startlink[1]{}%
\providecommand \@@endlink[0]{}%
\providecommand \url  [0]{\begingroup\@sanitize@url \@url }%
\providecommand \@url [1]{\endgroup\@href {#1}{\urlprefix }}%
\providecommand \urlprefix  [0]{URL }%
\providecommand \Eprint [0]{\href }%
\providecommand \doibase [0]{http://dx.doi.org/}%
\providecommand \selectlanguage [0]{\@gobble}%
\providecommand \bibinfo  [0]{\@secondoftwo}%
\providecommand \bibfield  [0]{\@secondoftwo}%
\providecommand \translation [1]{[#1]}%
\providecommand \BibitemOpen [0]{}%
\providecommand \bibitemStop [0]{}%
\providecommand \bibitemNoStop [0]{.\EOS\space}%
\providecommand \EOS [0]{\spacefactor3000\relax}%
\providecommand \BibitemShut  [1]{\csname bibitem#1\endcsname}%
\let\auto@bib@innerbib\@empty
\bibitem [{\citenamefont {Kippelen}\ and\ \citenamefont
  {Brédas}(2009)}]{solar}%
  \BibitemOpen
  \bibfield  {author} {\bibinfo {author} {\bibfnamefont {B.}~\bibnamefont
  {Kippelen}}\ and\ \bibinfo {author} {\bibfnamefont {J.-L.}\ \bibnamefont
  {Brédas}},\ }\href {\doibase 10.1039/B812502N} {\bibfield  {journal}
  {\bibinfo  {journal} {Energy Environ. Sci.}\ }\textbf {\bibinfo {volume}
  {2}},\ \bibinfo {pages} {251} (\bibinfo {year} {2009})}\BibitemShut {NoStop}%
\bibitem [{\citenamefont {Kulkarni}\ \emph {et~al.}(2004)\citenamefont
  {Kulkarni}, \citenamefont {Tonzola}, \citenamefont {Babel},\ and\
  \citenamefont {Jenekhe}}]{oled1}%
  \BibitemOpen
  \bibfield  {author} {\bibinfo {author} {\bibfnamefont {A.~P.}\ \bibnamefont
  {Kulkarni}}, \bibinfo {author} {\bibfnamefont {C.~J.}\ \bibnamefont
  {Tonzola}}, \bibinfo {author} {\bibfnamefont {A.}~\bibnamefont {Babel}}, \
  and\ \bibinfo {author} {\bibfnamefont {S.~A.}\ \bibnamefont {Jenekhe}},\
  }\href {\doibase 10.1021/cm049473l} {\bibfield  {journal} {\bibinfo
  {journal} {Chemistry of Materials}\ }\textbf {\bibinfo {volume} {16}},\
  \bibinfo {pages} {4556} (\bibinfo {year} {2004})},\ \Eprint
  {http://arxiv.org/abs/https://doi.org/10.1021/cm049473l}
  {https://doi.org/10.1021/cm049473l} \BibitemShut {NoStop}%
\bibitem [{\citenamefont {Wong}\ and\ \citenamefont
  {Zysman-Colman}(2017)}]{oled2}%
  \BibitemOpen
  \bibfield  {author} {\bibinfo {author} {\bibfnamefont {M.~Y.}\ \bibnamefont
  {Wong}}\ and\ \bibinfo {author} {\bibfnamefont {E.}~\bibnamefont
  {Zysman-Colman}},\ }\href {\doibase 10.1002/adma.201605444} {\bibfield
  {journal} {\bibinfo  {journal} {Advanced Materials}\ }\textbf {\bibinfo
  {volume} {29}},\ \bibinfo {pages} {1605444} (\bibinfo {year} {2017})},\
  \Eprint
  {http://arxiv.org/abs/https://onlinelibrary.wiley.com/doi/pdf/10.1002/adma.201605444}
  {https://onlinelibrary.wiley.com/doi/pdf/10.1002/adma.201605444} \BibitemShut
  {NoStop}%
\bibitem [{\citenamefont {Prasad}(1991)}]{Prasad1991}%
  \BibitemOpen
  \bibfield  {author} {\bibinfo {author} {\bibfnamefont {W.}~\bibnamefont
  {Prasad}},\ }\href
  {https://www.ebook.de/de/product/3597383/prasad_williams_nonlinear_optical.html}
  {\emph {\bibinfo {title} {Nonlinear Optical}}}\ (\bibinfo  {publisher} {John
  Wiley \& Sons},\ \bibinfo {year} {1991})\BibitemShut {NoStop}%
\bibitem [{zys(1994)}]{zyss1994}%
  \BibitemOpen
  \href {\doibase 10.1016/c2009-0-21331-6} {\emph {\bibinfo {title} {Molecular
  Nonlinear Optics}}}\ (\bibinfo  {publisher} {Elsevier},\ \bibinfo {year}
  {1994})\BibitemShut {NoStop}%
\bibitem [{\citenamefont {Terenziani}\ \emph
  {et~al.}(2008{\natexlab{a}})\citenamefont {Terenziani}, \citenamefont
  {Katan}, \citenamefont {Badaeva}, \citenamefont {Tretiak},\ and\
  \citenamefont {Blanchard-Desce}}]{TerenzianiReview2008}%
  \BibitemOpen
  \bibfield  {author} {\bibinfo {author} {\bibfnamefont {F.}~\bibnamefont
  {Terenziani}}, \bibinfo {author} {\bibfnamefont {C.}~\bibnamefont {Katan}},
  \bibinfo {author} {\bibfnamefont {E.}~\bibnamefont {Badaeva}}, \bibinfo
  {author} {\bibfnamefont {S.}~\bibnamefont {Tretiak}}, \ and\ \bibinfo
  {author} {\bibfnamefont {M.}~\bibnamefont {Blanchard-Desce}},\ }\href
  {\doibase 10.1002/adma.200800402} {\bibfield  {journal} {\bibinfo  {journal}
  {Advanced Materials}\ }\textbf {\bibinfo {volume} {20}},\ \bibinfo {pages}
  {4641} (\bibinfo {year} {2008}{\natexlab{a}})}\BibitemShut {NoStop}%
\bibitem [{\citenamefont {He}\ \emph {et~al.}(2008)\citenamefont {He},
  \citenamefont {Tan}, \citenamefont {Zheng},\ and\ \citenamefont
  {Prasad}}]{prasad2008}%
  \BibitemOpen
  \bibfield  {author} {\bibinfo {author} {\bibfnamefont {G.~S.}\ \bibnamefont
  {He}}, \bibinfo {author} {\bibfnamefont {L.-S.}\ \bibnamefont {Tan}},
  \bibinfo {author} {\bibfnamefont {Q.}~\bibnamefont {Zheng}}, \ and\ \bibinfo
  {author} {\bibfnamefont {P.~N.}\ \bibnamefont {Prasad}},\ }\href {\doibase
  10.1021/cr050054x} {\bibfield  {journal} {\bibinfo  {journal} {Chemical
  Reviews}\ }\textbf {\bibinfo {volume} {108}},\ \bibinfo {pages} {1245}
  (\bibinfo {year} {2008})}\BibitemShut {NoStop}%
\bibitem [{\citenamefont {Kovalenko}\ \emph {et~al.}(2003)\citenamefont
  {Kovalenko}, \citenamefont {Pérez~Lustres}, \citenamefont {Ernsting},\ and\
  \citenamefont {Rettig}}]{photoinduced}%
  \BibitemOpen
  \bibfield  {author} {\bibinfo {author} {\bibfnamefont {S.~A.}\ \bibnamefont
  {Kovalenko}}, \bibinfo {author} {\bibfnamefont {J.~L.}\ \bibnamefont
  {Pérez~Lustres}}, \bibinfo {author} {\bibfnamefont {N.~P.}\ \bibnamefont
  {Ernsting}}, \ and\ \bibinfo {author} {\bibfnamefont {W.}~\bibnamefont
  {Rettig}},\ }\href {\doibase 10.1021/jp026802t} {\bibfield  {journal}
  {\bibinfo  {journal} {The Journal of Physical Chemistry A}\ }\textbf
  {\bibinfo {volume} {107}},\ \bibinfo {pages} {10228} (\bibinfo {year}
  {2003})},\ \Eprint {http://arxiv.org/abs/https://doi.org/10.1021/jp026802t}
  {https://doi.org/10.1021/jp026802t} \BibitemShut {NoStop}%
\bibitem [{\citenamefont {Campioli}\ \emph {et~al.}(2019)\citenamefont
  {Campioli}, \citenamefont {Sanyal}, \citenamefont {Marcelli}, \citenamefont
  {Di~Donato}, \citenamefont {Blanchard-Desce}, \citenamefont {Mongin},
  \citenamefont {Painelli},\ and\ \citenamefont {Terenziani}}]{cv07a}%
  \BibitemOpen
  \bibfield  {author} {\bibinfo {author} {\bibfnamefont {E.}~\bibnamefont
  {Campioli}}, \bibinfo {author} {\bibfnamefont {S.}~\bibnamefont {Sanyal}},
  \bibinfo {author} {\bibfnamefont {A.}~\bibnamefont {Marcelli}}, \bibinfo
  {author} {\bibfnamefont {M.}~\bibnamefont {Di~Donato}}, \bibinfo {author}
  {\bibfnamefont {M.}~\bibnamefont {Blanchard-Desce}}, \bibinfo {author}
  {\bibfnamefont {O.}~\bibnamefont {Mongin}}, \bibinfo {author} {\bibfnamefont
  {A.}~\bibnamefont {Painelli}}, \ and\ \bibinfo {author} {\bibfnamefont
  {F.}~\bibnamefont {Terenziani}},\ }\href {\doibase 10.1002/cphc.201900703}
  {\bibfield  {journal} {\bibinfo  {journal} {ChemPhysChem}\ }\textbf {\bibinfo
  {volume} {20}},\ \bibinfo {pages} {2860} (\bibinfo {year} {2019})},\ \Eprint
  {http://arxiv.org/abs/https://chemistry-europe.onlinelibrary.wiley.com/doi/pdf/10.1002/cphc.201900703}
  {https://chemistry-europe.onlinelibrary.wiley.com/doi/pdf/10.1002/cphc.201900703}
  \BibitemShut {NoStop}%
\bibitem [{\citenamefont {Terenziani}\ \emph {et~al.}(2007)\citenamefont
  {Terenziani}, \citenamefont {D'Avino},\ and\ \citenamefont
  {Painelli}}]{review2007}%
  \BibitemOpen
  \bibfield  {author} {\bibinfo {author} {\bibfnamefont {F.}~\bibnamefont
  {Terenziani}}, \bibinfo {author} {\bibfnamefont {G.}~\bibnamefont {D'Avino}},
  \ and\ \bibinfo {author} {\bibfnamefont {A.}~\bibnamefont {Painelli}},\
  }\href {\doibase 10.1002/cphc.200700368} {\bibfield  {journal} {\bibinfo
  {journal} {ChemPhysChem}\ }\textbf {\bibinfo {volume} {8}},\ \bibinfo {pages}
  {2433} (\bibinfo {year} {2007})},\ \Eprint
  {http://arxiv.org/abs/https://chemistry-europe.onlinelibrary.wiley.com/doi/pdf/10.1002/cphc.200700368}
  {https://chemistry-europe.onlinelibrary.wiley.com/doi/pdf/10.1002/cphc.200700368}
  \BibitemShut {NoStop}%
\bibitem [{\citenamefont {Terenziani}\ and\ \citenamefont
  {Painelli}(2003)}]{prb2003}%
  \BibitemOpen
  \bibfield  {author} {\bibinfo {author} {\bibfnamefont {F.}~\bibnamefont
  {Terenziani}}\ and\ \bibinfo {author} {\bibfnamefont {A.}~\bibnamefont
  {Painelli}},\ }\href {\doibase 10.1103/PhysRevB.68.165405} {\bibfield
  {journal} {\bibinfo  {journal} {Phys. Rev. B}\ }\textbf {\bibinfo {volume}
  {68}},\ \bibinfo {pages} {165405} (\bibinfo {year} {2003})}\BibitemShut
  {NoStop}%
\bibitem [{\citenamefont {Painelli}\ and\ \citenamefont
  {Terenziani}(2003)}]{jacs2003}%
  \BibitemOpen
  \bibfield  {author} {\bibinfo {author} {\bibfnamefont {A.}~\bibnamefont
  {Painelli}}\ and\ \bibinfo {author} {\bibfnamefont {F.}~\bibnamefont
  {Terenziani}},\ }\href {\doibase 10.1021/ja034155t} {\bibfield  {journal}
  {\bibinfo  {journal} {Journal of the American Chemical Society}\ }\textbf
  {\bibinfo {volume} {125}},\ \bibinfo {pages} {5624} (\bibinfo {year}
  {2003})},\ \bibinfo {note} {pMID: 12733888},\ \Eprint
  {http://arxiv.org/abs/https://doi.org/10.1021/ja034155t}
  {https://doi.org/10.1021/ja034155t} \BibitemShut {NoStop}%
\bibitem [{\citenamefont {Sanyal}\ \emph {et~al.}(2016)\citenamefont {Sanyal},
  \citenamefont {Painelli}, \citenamefont {Pati}, \citenamefont {Terenziani},\
  and\ \citenamefont {Sissa}}]{curcumine}%
  \BibitemOpen
  \bibfield  {author} {\bibinfo {author} {\bibfnamefont {S.}~\bibnamefont
  {Sanyal}}, \bibinfo {author} {\bibfnamefont {A.}~\bibnamefont {Painelli}},
  \bibinfo {author} {\bibfnamefont {S.~K.}\ \bibnamefont {Pati}}, \bibinfo
  {author} {\bibfnamefont {F.}~\bibnamefont {Terenziani}}, \ and\ \bibinfo
  {author} {\bibfnamefont {C.}~\bibnamefont {Sissa}},\ }\href {\doibase
  10.1039/C6CP05153G} {\bibfield  {journal} {\bibinfo  {journal} {Phys. Chem.
  Chem. Phys.}\ }\textbf {\bibinfo {volume} {18}},\ \bibinfo {pages} {28198}
  (\bibinfo {year} {2016})}\BibitemShut {NoStop}%
\bibitem [{\citenamefont {Sanyal}\ \emph {et~al.}(2017)\citenamefont {Sanyal},
  \citenamefont {Sissa}, \citenamefont {Terenziani}, \citenamefont {Pati},\
  and\ \citenamefont {Painelli}}]{dans2017}%
  \BibitemOpen
  \bibfield  {author} {\bibinfo {author} {\bibfnamefont {S.}~\bibnamefont
  {Sanyal}}, \bibinfo {author} {\bibfnamefont {C.}~\bibnamefont {Sissa}},
  \bibinfo {author} {\bibfnamefont {F.}~\bibnamefont {Terenziani}}, \bibinfo
  {author} {\bibfnamefont {S.~K.}\ \bibnamefont {Pati}}, \ and\ \bibinfo
  {author} {\bibfnamefont {A.}~\bibnamefont {Painelli}},\ }\href {\doibase
  10.1039/C7CP04732K} {\bibfield  {journal} {\bibinfo  {journal} {Phys. Chem.
  Chem. Phys.}\ }\textbf {\bibinfo {volume} {19}},\ \bibinfo {pages} {24979}
  (\bibinfo {year} {2017})}\BibitemShut {NoStop}%
\bibitem [{\citenamefont {Bardi}\ \emph {et~al.}(2017)\citenamefont {Bardi},
  \citenamefont {Dall’Agnese}, \citenamefont {Moineau-Chane~Ching},
  \citenamefont {Painelli},\ and\ \citenamefont {Terenziani}}]{brunella1}%
  \BibitemOpen
  \bibfield  {author} {\bibinfo {author} {\bibfnamefont {B.}~\bibnamefont
  {Bardi}}, \bibinfo {author} {\bibfnamefont {C.}~\bibnamefont
  {Dall’Agnese}}, \bibinfo {author} {\bibfnamefont {K.~I.}\ \bibnamefont
  {Moineau-Chane~Ching}}, \bibinfo {author} {\bibfnamefont {A.}~\bibnamefont
  {Painelli}}, \ and\ \bibinfo {author} {\bibfnamefont {F.}~\bibnamefont
  {Terenziani}},\ }\href {\doibase 10.1021/acs.jpcc.7b04647} {\bibfield
  {journal} {\bibinfo  {journal} {The Journal of Physical Chemistry C}\
  }\textbf {\bibinfo {volume} {121}},\ \bibinfo {pages} {17466} (\bibinfo
  {year} {2017})},\ \Eprint
  {http://arxiv.org/abs/https://doi.org/10.1021/acs.jpcc.7b04647}
  {https://doi.org/10.1021/acs.jpcc.7b04647} \BibitemShut {NoStop}%
\bibitem [{\citenamefont {Bardi}\ \emph {et~al.}(2018)\citenamefont {Bardi},
  \citenamefont {Dall'Agnese}, \citenamefont {Tassé}, \citenamefont {Ladeira},
  \citenamefont {Painelli}, \citenamefont {Moineau-Chane-Ching},\ and\
  \citenamefont {Terenziani}}]{brunella2}%
  \BibitemOpen
  \bibfield  {author} {\bibinfo {author} {\bibfnamefont {B.}~\bibnamefont
  {Bardi}}, \bibinfo {author} {\bibfnamefont {C.}~\bibnamefont {Dall'Agnese}},
  \bibinfo {author} {\bibfnamefont {M.}~\bibnamefont {Tassé}}, \bibinfo
  {author} {\bibfnamefont {S.}~\bibnamefont {Ladeira}}, \bibinfo {author}
  {\bibfnamefont {A.}~\bibnamefont {Painelli}}, \bibinfo {author}
  {\bibfnamefont {K.~I.}\ \bibnamefont {Moineau-Chane-Ching}}, \ and\ \bibinfo
  {author} {\bibfnamefont {F.}~\bibnamefont {Terenziani}},\ }\href {\doibase
  10.1002/cptc.201800145} {\bibfield  {journal} {\bibinfo  {journal}
  {ChemPhotoChem}\ }\textbf {\bibinfo {volume} {2}},\ \bibinfo {pages} {1027}
  (\bibinfo {year} {2018})},\ \Eprint
  {http://arxiv.org/abs/https://chemistry-europe.onlinelibrary.wiley.com/doi/pdf/10.1002/cptc.201800145}
  {https://chemistry-europe.onlinelibrary.wiley.com/doi/pdf/10.1002/cptc.201800145}
  \BibitemShut {NoStop}%
\bibitem [{\citenamefont {Zhong}\ \emph {et~al.}(2020)\citenamefont {Zhong},
  \citenamefont {Bialas},\ and\ \citenamefont {Spano}}]{spano1}%
  \BibitemOpen
  \bibfield  {author} {\bibinfo {author} {\bibfnamefont {C.}~\bibnamefont
  {Zhong}}, \bibinfo {author} {\bibfnamefont {D.}~\bibnamefont {Bialas}}, \
  and\ \bibinfo {author} {\bibfnamefont {F.~C.}\ \bibnamefont {Spano}},\ }\href
  {\doibase 10.1021/acs.jpcc.9b09368} {\bibfield  {journal} {\bibinfo
  {journal} {The Journal of Physical Chemistry C}\ }\textbf {\bibinfo {volume}
  {124}},\ \bibinfo {pages} {2146} (\bibinfo {year} {2020})},\ \Eprint
  {http://arxiv.org/abs/https://doi.org/10.1021/acs.jpcc.9b09368}
  {https://doi.org/10.1021/acs.jpcc.9b09368} \BibitemShut {NoStop}%
\bibitem [{\citenamefont {Zheng}\ \emph {et~al.}(2019)\citenamefont {Zheng},
  \citenamefont {Zhong}, \citenamefont {Collison},\ and\ \citenamefont
  {Spano}}]{spano2}%
  \BibitemOpen
  \bibfield  {author} {\bibinfo {author} {\bibfnamefont {C.}~\bibnamefont
  {Zheng}}, \bibinfo {author} {\bibfnamefont {C.}~\bibnamefont {Zhong}},
  \bibinfo {author} {\bibfnamefont {C.~J.}\ \bibnamefont {Collison}}, \ and\
  \bibinfo {author} {\bibfnamefont {F.~C.}\ \bibnamefont {Spano}},\ }\href
  {\doibase 10.1021/acs.jpcc.8b11416} {\bibfield  {journal} {\bibinfo
  {journal} {The Journal of Physical Chemistry C}\ }\textbf {\bibinfo {volume}
  {123}},\ \bibinfo {pages} {3203} (\bibinfo {year} {2019})},\ \Eprint
  {http://arxiv.org/abs/https://doi.org/10.1021/acs.jpcc.8b11416}
  {https://doi.org/10.1021/acs.jpcc.8b11416} \BibitemShut {NoStop}%
\bibitem [{\citenamefont {Liptay}(1969)}]{liptay}%
  \BibitemOpen
  \bibfield  {author} {\bibinfo {author} {\bibfnamefont {W.}~\bibnamefont
  {Liptay}},\ }\href {\doibase 10.1002/anie.196901771} {\bibfield  {journal}
  {\bibinfo  {journal} {Angewandte Chemie International Edition in English}\
  }\textbf {\bibinfo {volume} {8}},\ \bibinfo {pages} {177} (\bibinfo {year}
  {1969})},\ \Eprint
  {http://arxiv.org/abs/https://onlinelibrary.wiley.com/doi/pdf/10.1002/anie.196901771}
  {https://onlinelibrary.wiley.com/doi/pdf/10.1002/anie.196901771} \BibitemShut
  {NoStop}%
\bibitem [{\citenamefont {Reichardt}(1994)}]{reichardt}%
  \BibitemOpen
  \bibfield  {author} {\bibinfo {author} {\bibfnamefont {C.}~\bibnamefont
  {Reichardt}},\ }\href {\doibase 10.1021/cr00032a005} {\bibfield  {journal}
  {\bibinfo  {journal} {Chemical Reviews}\ }\textbf {\bibinfo {volume} {94}},\
  \bibinfo {pages} {2319} (\bibinfo {year} {1994})},\ \Eprint
  {http://arxiv.org/abs/https://doi.org/10.1021/cr00032a005}
  {https://doi.org/10.1021/cr00032a005} \BibitemShut {NoStop}%
\bibitem [{\citenamefont {Terenziani}\ \emph {et~al.}(2006)\citenamefont
  {Terenziani}, \citenamefont {Painelli}, \citenamefont {Katan}, \citenamefont
  {Charlot},\ and\ \citenamefont {Blanchard-Desce}}]{jacs2006}%
  \BibitemOpen
  \bibfield  {author} {\bibinfo {author} {\bibfnamefont {F.}~\bibnamefont
  {Terenziani}}, \bibinfo {author} {\bibfnamefont {A.}~\bibnamefont
  {Painelli}}, \bibinfo {author} {\bibfnamefont {C.}~\bibnamefont {Katan}},
  \bibinfo {author} {\bibfnamefont {M.}~\bibnamefont {Charlot}}, \ and\
  \bibinfo {author} {\bibfnamefont {M.}~\bibnamefont {Blanchard-Desce}},\
  }\href {\doibase 10.1021/ja064521j} {\bibfield  {journal} {\bibinfo
  {journal} {Journal of the American Chemical Society}\ }\textbf {\bibinfo
  {volume} {128}},\ \bibinfo {pages} {15742} (\bibinfo {year} {2006})},\
  \bibinfo {note} {pMID: 17147384},\ \Eprint
  {http://arxiv.org/abs/https://doi.org/10.1021/ja064521j}
  {https://doi.org/10.1021/ja064521j} \BibitemShut {NoStop}%
\bibitem [{\citenamefont {Campo}\ \emph {et~al.}(2010)\citenamefont {Campo},
  \citenamefont {Painelli}, \citenamefont {Terenziani}, \citenamefont
  {Van~Regemorter}, \citenamefont {Beljonne}, \citenamefont {Goovaerts},\ and\
  \citenamefont {Wenseleers}}]{jacs2010}%
  \BibitemOpen
  \bibfield  {author} {\bibinfo {author} {\bibfnamefont {J.}~\bibnamefont
  {Campo}}, \bibinfo {author} {\bibfnamefont {A.}~\bibnamefont {Painelli}},
  \bibinfo {author} {\bibfnamefont {F.}~\bibnamefont {Terenziani}}, \bibinfo
  {author} {\bibfnamefont {T.}~\bibnamefont {Van~Regemorter}}, \bibinfo
  {author} {\bibfnamefont {D.}~\bibnamefont {Beljonne}}, \bibinfo {author}
  {\bibfnamefont {E.}~\bibnamefont {Goovaerts}}, \ and\ \bibinfo {author}
  {\bibfnamefont {W.}~\bibnamefont {Wenseleers}},\ }\href {\doibase
  10.1021/ja105600t} {\bibfield  {journal} {\bibinfo  {journal} {Journal of the
  American Chemical Society}\ }\textbf {\bibinfo {volume} {132}},\ \bibinfo
  {pages} {16467} (\bibinfo {year} {2010})},\ \bibinfo {note} {pMID:
  21033705},\ \Eprint {http://arxiv.org/abs/https://doi.org/10.1021/ja105600t}
  {https://doi.org/10.1021/ja105600t} \BibitemShut {NoStop}%
\bibitem [{\citenamefont {Terenziani}\ \emph {et~al.}(2010)\citenamefont
  {Terenziani}, \citenamefont {Przhonska}, \citenamefont {Webster},
  \citenamefont {Padilha}, \citenamefont {Slominsky}, \citenamefont
  {Davydenko}, \citenamefont {Gerasov}, \citenamefont {Kovtun}, \citenamefont
  {Shandura}, \citenamefont {Kachkovski}, \citenamefont {Hagan}, \citenamefont
  {Van~Stryland},\ and\ \citenamefont {Painelli}}]{jpcl2010}%
  \BibitemOpen
  \bibfield  {author} {\bibinfo {author} {\bibfnamefont {F.}~\bibnamefont
  {Terenziani}}, \bibinfo {author} {\bibfnamefont {O.~V.}\ \bibnamefont
  {Przhonska}}, \bibinfo {author} {\bibfnamefont {S.}~\bibnamefont {Webster}},
  \bibinfo {author} {\bibfnamefont {L.~A.}\ \bibnamefont {Padilha}}, \bibinfo
  {author} {\bibfnamefont {Y.~L.}\ \bibnamefont {Slominsky}}, \bibinfo {author}
  {\bibfnamefont {I.~G.}\ \bibnamefont {Davydenko}}, \bibinfo {author}
  {\bibfnamefont {A.~O.}\ \bibnamefont {Gerasov}}, \bibinfo {author}
  {\bibfnamefont {Y.~P.}\ \bibnamefont {Kovtun}}, \bibinfo {author}
  {\bibfnamefont {M.~P.}\ \bibnamefont {Shandura}}, \bibinfo {author}
  {\bibfnamefont {A.~D.}\ \bibnamefont {Kachkovski}}, \bibinfo {author}
  {\bibfnamefont {D.~J.}\ \bibnamefont {Hagan}}, \bibinfo {author}
  {\bibfnamefont {E.~W.}\ \bibnamefont {Van~Stryland}}, \ and\ \bibinfo
  {author} {\bibfnamefont {A.}~\bibnamefont {Painelli}},\ }\href {\doibase
  10.1021/jz100430x} {\bibfield  {journal} {\bibinfo  {journal} {The Journal of
  Physical Chemistry Letters}\ }\textbf {\bibinfo {volume} {1}},\ \bibinfo
  {pages} {1800} (\bibinfo {year} {2010})},\ \Eprint
  {http://arxiv.org/abs/https://doi.org/10.1021/jz100430x}
  {https://doi.org/10.1021/jz100430x} \BibitemShut {NoStop}%
\bibitem [{\citenamefont {Hu}\ \emph {et~al.}(2013)\citenamefont {Hu},
  \citenamefont {Przhonska}, \citenamefont {Terenziani}, \citenamefont
  {Painelli}, \citenamefont {Fishman}, \citenamefont {Ensley}, \citenamefont
  {Reichert}, \citenamefont {Webster}, \citenamefont {Bricks}, \citenamefont
  {Kachkovski}, \citenamefont {Hagan},\ and\ \citenamefont
  {Van~Stryland}}]{vanstryland}%
  \BibitemOpen
  \bibfield  {author} {\bibinfo {author} {\bibfnamefont {H.}~\bibnamefont
  {Hu}}, \bibinfo {author} {\bibfnamefont {O.~V.}\ \bibnamefont {Przhonska}},
  \bibinfo {author} {\bibfnamefont {F.}~\bibnamefont {Terenziani}}, \bibinfo
  {author} {\bibfnamefont {A.}~\bibnamefont {Painelli}}, \bibinfo {author}
  {\bibfnamefont {D.}~\bibnamefont {Fishman}}, \bibinfo {author} {\bibfnamefont
  {T.~R.}\ \bibnamefont {Ensley}}, \bibinfo {author} {\bibfnamefont
  {M.}~\bibnamefont {Reichert}}, \bibinfo {author} {\bibfnamefont
  {S.}~\bibnamefont {Webster}}, \bibinfo {author} {\bibfnamefont {J.~L.}\
  \bibnamefont {Bricks}}, \bibinfo {author} {\bibfnamefont {A.~D.}\
  \bibnamefont {Kachkovski}}, \bibinfo {author} {\bibfnamefont {D.~J.}\
  \bibnamefont {Hagan}}, \ and\ \bibinfo {author} {\bibfnamefont {E.~W.}\
  \bibnamefont {Van~Stryland}},\ }\href {\doibase 10.1039/C3CP50811K}
  {\bibfield  {journal} {\bibinfo  {journal} {Phys. Chem. Chem. Phys.}\
  }\textbf {\bibinfo {volume} {15}},\ \bibinfo {pages} {7666} (\bibinfo {year}
  {2013})}\BibitemShut {NoStop}%
\bibitem [{\citenamefont {Painelli}\ and\ \citenamefont
  {Terenziani}(1999)}]{chemphyslett1999}%
  \BibitemOpen
  \bibfield  {author} {\bibinfo {author} {\bibfnamefont {A.}~\bibnamefont
  {Painelli}}\ and\ \bibinfo {author} {\bibfnamefont {F.}~\bibnamefont
  {Terenziani}},\ }\href {\doibase
  https://doi.org/10.1016/S0009-2614(99)00960-4} {\bibfield  {journal}
  {\bibinfo  {journal} {Chemical Physics Letters}\ }\textbf {\bibinfo {volume}
  {312}},\ \bibinfo {pages} {211 } (\bibinfo {year} {1999})}\BibitemShut
  {NoStop}%
\bibitem [{\citenamefont {Boldrini}\ \emph {et~al.}(2002)\citenamefont
  {Boldrini}, \citenamefont {Cavalli}, \citenamefont {Painelli},\ and\
  \citenamefont {Terenziani}}]{baba}%
  \BibitemOpen
  \bibfield  {author} {\bibinfo {author} {\bibfnamefont {B.}~\bibnamefont
  {Boldrini}}, \bibinfo {author} {\bibfnamefont {E.}~\bibnamefont {Cavalli}},
  \bibinfo {author} {\bibfnamefont {A.}~\bibnamefont {Painelli}}, \ and\
  \bibinfo {author} {\bibfnamefont {F.}~\bibnamefont {Terenziani}},\
  }\href@noop {} {\bibfield  {journal} {\bibinfo  {journal} {J. Phys. Chem. A}\
  }\textbf {\bibinfo {volume} {106}},\ \bibinfo {pages} {6286} (\bibinfo {year}
  {2002})}\BibitemShut {NoStop}%
\bibitem [{\citenamefont {Terenziani}\ \emph
  {et~al.}(2008{\natexlab{b}})\citenamefont {Terenziani}, \citenamefont
  {Sissa},\ and\ \citenamefont {Painelli}}]{octupolar}%
  \BibitemOpen
  \bibfield  {author} {\bibinfo {author} {\bibfnamefont {F.}~\bibnamefont
  {Terenziani}}, \bibinfo {author} {\bibfnamefont {C.}~\bibnamefont {Sissa}}, \
  and\ \bibinfo {author} {\bibfnamefont {A.}~\bibnamefont {Painelli}},\ }\href
  {\doibase 10.1021/jp710241g} {\bibfield  {journal} {\bibinfo  {journal} {The
  Journal of Physical Chemistry B}\ }\textbf {\bibinfo {volume} {112}},\
  \bibinfo {pages} {5079} (\bibinfo {year} {2008}{\natexlab{b}})},\ \bibinfo
  {note} {pMID: 18376886},\ \Eprint
  {http://arxiv.org/abs/https://doi.org/10.1021/jp710241g}
  {https://doi.org/10.1021/jp710241g} \BibitemShut {NoStop}%
\bibitem [{\citenamefont {Di~Bella}\ \emph {et~al.}(1994)\citenamefont
  {Di~Bella}, \citenamefont {Marks},\ and\ \citenamefont {Ratner}}]{dibella}%
  \BibitemOpen
  \bibfield  {author} {\bibinfo {author} {\bibfnamefont {S.}~\bibnamefont
  {Di~Bella}}, \bibinfo {author} {\bibfnamefont {T.~J.}\ \bibnamefont {Marks}},
  \ and\ \bibinfo {author} {\bibfnamefont {M.~A.}\ \bibnamefont {Ratner}},\
  }\href@noop {} {\bibfield  {journal} {\bibinfo  {journal} {J. Am. Chem.
  Soc.}\ }\textbf {\bibinfo {volume} {116}},\ \bibinfo {pages} {4440} (\bibinfo
  {year} {1994})}\BibitemShut {NoStop}%
\bibitem [{\citenamefont {Painelli}(1999)}]{cp99}%
  \BibitemOpen
  \bibfield  {author} {\bibinfo {author} {\bibfnamefont {A.}~\bibnamefont
  {Painelli}},\ }\href@noop {} {\bibfield  {journal} {\bibinfo  {journal}
  {Chem. Phys.}\ }\textbf {\bibinfo {volume} {245}},\ \bibinfo {pages} {185}
  (\bibinfo {year} {1999})}\BibitemShut {NoStop}%
\bibitem [{\citenamefont {Phan~Huu}\ \emph {et~al.}(2020)\citenamefont
  {Phan~Huu}, \citenamefont {Dhali}, \citenamefont {Pieroni}, \citenamefont
  {Di~Maiolo}, \citenamefont {Sissa}, \citenamefont {Terenziani},\ and\
  \citenamefont {Painelli}}]{prl2020}%
  \BibitemOpen
  \bibfield  {author} {\bibinfo {author} {\bibfnamefont {D.~K.~A.}\
  \bibnamefont {Phan~Huu}}, \bibinfo {author} {\bibfnamefont {R.}~\bibnamefont
  {Dhali}}, \bibinfo {author} {\bibfnamefont {C.}~\bibnamefont {Pieroni}},
  \bibinfo {author} {\bibfnamefont {F.}~\bibnamefont {Di~Maiolo}}, \bibinfo
  {author} {\bibfnamefont {C.}~\bibnamefont {Sissa}}, \bibinfo {author}
  {\bibfnamefont {F.}~\bibnamefont {Terenziani}}, \ and\ \bibinfo {author}
  {\bibfnamefont {A.}~\bibnamefont {Painelli}},\ }\href {\doibase
  10.1103/PhysRevLett.124.107401} {\bibfield  {journal} {\bibinfo  {journal}
  {Phys. Rev. Lett.}\ }\textbf {\bibinfo {volume} {124}},\ \bibinfo {pages}
  {107401} (\bibinfo {year} {2020})}\BibitemShut {NoStop}%
\bibitem [{\citenamefont {Tomasi}\ \emph {et~al.}(2005)\citenamefont {Tomasi},
  \citenamefont {Mennucci},\ and\ \citenamefont {Cammi}}]{tomasichemrev}%
  \BibitemOpen
  \bibfield  {author} {\bibinfo {author} {\bibfnamefont {J.}~\bibnamefont
  {Tomasi}}, \bibinfo {author} {\bibfnamefont {B.}~\bibnamefont {Mennucci}}, \
  and\ \bibinfo {author} {\bibfnamefont {R.}~\bibnamefont {Cammi}},\
  }\href@noop {} {\bibfield  {journal} {\bibinfo  {journal} {Chem. Rev.}\
  }\textbf {\bibinfo {volume} {105}},\ \bibinfo {pages} {2999} (\bibinfo {year}
  {2005})}\BibitemShut {NoStop}%
\bibitem [{\citenamefont {Marenich}\ \emph {et~al.}(2011)\citenamefont
  {Marenich}, \citenamefont {Cramer}, \citenamefont {Truhlar}, \citenamefont
  {Guido}, \citenamefont {Mennucci}, \citenamefont {Scalmani},\ and\
  \citenamefont {Frisch}}]{chemsci2011}%
  \BibitemOpen
  \bibfield  {author} {\bibinfo {author} {\bibfnamefont {A.~V.}\ \bibnamefont
  {Marenich}}, \bibinfo {author} {\bibfnamefont {C.~J.}\ \bibnamefont
  {Cramer}}, \bibinfo {author} {\bibfnamefont {D.~G.}\ \bibnamefont {Truhlar}},
  \bibinfo {author} {\bibfnamefont {C.~A.}\ \bibnamefont {Guido}}, \bibinfo
  {author} {\bibfnamefont {B.}~\bibnamefont {Mennucci}}, \bibinfo {author}
  {\bibfnamefont {G.}~\bibnamefont {Scalmani}}, \ and\ \bibinfo {author}
  {\bibfnamefont {M.~J.}\ \bibnamefont {Frisch}},\ }\href@noop {} {\bibfield
  {journal} {\bibinfo  {journal} {Chem. Sci.}\ }\textbf {\bibinfo {volume}
  {2}},\ \bibinfo {pages} {2143} (\bibinfo {year} {2011})}\BibitemShut
  {NoStop}%
\bibitem [{\citenamefont {Lunkenheimer}\ and\ \citenamefont
  {K\"ohn}(2013)}]{lunkenheimer2012}%
  \BibitemOpen
  \bibfield  {author} {\bibinfo {author} {\bibfnamefont {B.}~\bibnamefont
  {Lunkenheimer}}\ and\ \bibinfo {author} {\bibnamefont {K\"ohn}},\ }\href@noop
  {} {\bibfield  {journal} {\bibinfo  {journal} {J. Chem. Theory Comput.}\
  }\textbf {\bibinfo {volume} {9}},\ \bibinfo {pages} {977} (\bibinfo {year}
  {2013})}\BibitemShut {NoStop}%
\bibitem [{\citenamefont {Guido}\ and\ \citenamefont
  {Caprasecca}(2019{\natexlab{a}})}]{ciroguido}%
  \BibitemOpen
  \bibfield  {author} {\bibinfo {author} {\bibfnamefont {C.~A.}\ \bibnamefont
  {Guido}}\ and\ \bibinfo {author} {\bibfnamefont {S.}~\bibnamefont
  {Caprasecca}},\ }\href {\doibase 10.1002/qua.25711} {\bibfield  {journal}
  {\bibinfo  {journal} {Int. J. Quantum Chem.}\ }\textbf {\bibinfo {volume}
  {119}},\ \bibinfo {pages} {e25711} (\bibinfo {year}
  {2019}{\natexlab{a}})}\BibitemShut {NoStop}%
\bibitem [{\citenamefont {Li}\ \emph {et~al.}(2018)\citenamefont {Li},
  \citenamefont {D'Avino}, \citenamefont {Duchemin}, \citenamefont {Beljonne},\
  and\ \citenamefont {Blase}}]{gabriele2018}%
  \BibitemOpen
  \bibfield  {author} {\bibinfo {author} {\bibfnamefont {J.}~\bibnamefont
  {Li}}, \bibinfo {author} {\bibfnamefont {G.}~\bibnamefont {D'Avino}},
  \bibinfo {author} {\bibfnamefont {I.}~\bibnamefont {Duchemin}}, \bibinfo
  {author} {\bibfnamefont {D.}~\bibnamefont {Beljonne}}, \ and\ \bibinfo
  {author} {\bibfnamefont {X.}~\bibnamefont {Blase}},\ }\href@noop {}
  {\bibfield  {journal} {\bibinfo  {journal} {Phys. Rev. B}\ }\textbf {\bibinfo
  {volume} {97}},\ \bibinfo {pages} {035108} (\bibinfo {year}
  {2018})}\BibitemShut {NoStop}%
\bibitem [{\citenamefont {Vreven}\ and\ \citenamefont {Morokuma}(2006)}]{qmmm}%
  \BibitemOpen
  \bibfield  {author} {\bibinfo {author} {\bibfnamefont {T.}~\bibnamefont
  {Vreven}}\ and\ \bibinfo {author} {\bibfnamefont {K.}~\bibnamefont
  {Morokuma}}\ }(\bibinfo  {publisher} {Elsevier},\ \bibinfo {year} {2006})\
  pp.\ \bibinfo {pages} {35 -- 51}\BibitemShut {NoStop}%
\bibitem [{\citenamefont {McRae}(1957)}]{mcrae}%
  \BibitemOpen
  \bibfield  {author} {\bibinfo {author} {\bibfnamefont {E.~G.}\ \bibnamefont
  {McRae}},\ }\href@noop {} {\bibfield  {journal} {\bibinfo  {journal}
  {J.~Phys. Chem.}\ }\textbf {\bibinfo {volume} {61}},\ \bibinfo {pages} {1128}
  (\bibinfo {year} {1957})}\BibitemShut {NoStop}%
\bibitem [{\citenamefont {Feinberg}\ \emph {et~al.}(1990)\citenamefont
  {Feinberg}, \citenamefont {Ciuchi},\ and\ \citenamefont
  {De~Pasquale}}]{ciuchi}%
  \BibitemOpen
  \bibfield  {author} {\bibinfo {author} {\bibfnamefont {D.}~\bibnamefont
  {Feinberg}}, \bibinfo {author} {\bibfnamefont {S.}~\bibnamefont {Ciuchi}}, \
  and\ \bibinfo {author} {\bibfnamefont {F.}~\bibnamefont {De~Pasquale}},\
  }\href@noop {} {\bibfield  {journal} {\bibinfo  {journal} {Int. J. Modern
  Phys. B}\ }\textbf {\bibinfo {volume} {4}},\ \bibinfo {pages} {1317}
  (\bibinfo {year} {1990})}\BibitemShut {NoStop}%
\bibitem [{\citenamefont {Cammi}\ \emph {et~al.}(2005)\citenamefont {Cammi},
  \citenamefont {Corni}, \citenamefont {Mennucci},\ and\ \citenamefont
  {Tomasi}}]{cammi:lr-ss}%
  \BibitemOpen
  \bibfield  {author} {\bibinfo {author} {\bibfnamefont {R.}~\bibnamefont
  {Cammi}}, \bibinfo {author} {\bibfnamefont {S.}~\bibnamefont {Corni}},
  \bibinfo {author} {\bibfnamefont {B.}~\bibnamefont {Mennucci}}, \ and\
  \bibinfo {author} {\bibfnamefont {J.}~\bibnamefont {Tomasi}},\ }\href
  {\doibase 10.1063/1.1867373} {\bibfield  {journal} {\bibinfo  {journal} {The
  Journal of Chemical Physics}\ }\textbf {\bibinfo {volume} {122}},\ \bibinfo
  {pages} {104513} (\bibinfo {year} {2005})},\ \Eprint
  {http://arxiv.org/abs/https://doi.org/10.1063/1.1867373}
  {https://doi.org/10.1063/1.1867373} \BibitemShut {NoStop}%
\bibitem [{\citenamefont {Improta}\ \emph
  {et~al.}(2006{\natexlab{a}})\citenamefont {Improta}, \citenamefont {Barone},
  \citenamefont {Scalmani},\ and\ \citenamefont {Frisch}}]{improta}%
  \BibitemOpen
  \bibfield  {author} {\bibinfo {author} {\bibfnamefont {R.}~\bibnamefont
  {Improta}}, \bibinfo {author} {\bibfnamefont {V.}~\bibnamefont {Barone}},
  \bibinfo {author} {\bibfnamefont {G.}~\bibnamefont {Scalmani}}, \ and\
  \bibinfo {author} {\bibfnamefont {M.~J.}\ \bibnamefont {Frisch}},\ }\href
  {\doibase 10.1063/1.2222364} {\bibfield  {journal} {\bibinfo  {journal} {The
  Journal of Chemical Physics}\ }\textbf {\bibinfo {volume} {125}},\ \bibinfo
  {pages} {054103} (\bibinfo {year} {2006}{\natexlab{a}})},\ \Eprint
  {http://arxiv.org/abs/https://doi.org/10.1063/1.2222364}
  {https://doi.org/10.1063/1.2222364} \BibitemShut {NoStop}%
\bibitem [{\citenamefont {Guido}\ and\ \citenamefont
  {Caprasecca}(2019{\natexlab{b}})}]{caprasecca}%
  \BibitemOpen
  \bibfield  {author} {\bibinfo {author} {\bibfnamefont {C.~A.}\ \bibnamefont
  {Guido}}\ and\ \bibinfo {author} {\bibfnamefont {S.}~\bibnamefont
  {Caprasecca}},\ }\href {\doibase 10.1002/qua.25711} {\bibfield  {journal}
  {\bibinfo  {journal} {International Journal of Quantum Chemistry}\ }\textbf
  {\bibinfo {volume} {119}},\ \bibinfo {pages} {e25711} (\bibinfo {year}
  {2019}{\natexlab{b}})},\ \Eprint
  {http://arxiv.org/abs/https://onlinelibrary.wiley.com/doi/pdf/10.1002/qua.25711}
  {https://onlinelibrary.wiley.com/doi/pdf/10.1002/qua.25711} \BibitemShut
  {NoStop}%
\bibitem [{\citenamefont {Nakanotani}\ \emph {et~al.}(2014)\citenamefont
  {Nakanotani}, \citenamefont {Higuchi}, \citenamefont {Furukawa},
  \citenamefont {Masui}, \citenamefont {Morimoto}, \citenamefont {Numata},
  \citenamefont {Tanaka}, \citenamefont {Sagara}, \citenamefont {Yasuda},\ and\
  \citenamefont {Adachi}}]{adachi}%
  \BibitemOpen
  \bibfield  {author} {\bibinfo {author} {\bibfnamefont {H.}~\bibnamefont
  {Nakanotani}}, \bibinfo {author} {\bibfnamefont {T.}~\bibnamefont {Higuchi}},
  \bibinfo {author} {\bibfnamefont {T.}~\bibnamefont {Furukawa}}, \bibinfo
  {author} {\bibfnamefont {K.}~\bibnamefont {Masui}}, \bibinfo {author}
  {\bibfnamefont {K.}~\bibnamefont {Morimoto}}, \bibinfo {author}
  {\bibfnamefont {M.}~\bibnamefont {Numata}}, \bibinfo {author} {\bibfnamefont
  {H.}~\bibnamefont {Tanaka}}, \bibinfo {author} {\bibfnamefont
  {Y.}~\bibnamefont {Sagara}}, \bibinfo {author} {\bibfnamefont
  {T.}~\bibnamefont {Yasuda}}, \ and\ \bibinfo {author} {\bibfnamefont
  {C.}~\bibnamefont {Adachi}},\ }\href@noop {} {\bibfield  {journal} {\bibinfo
  {journal} {Nature Communications}\ }\textbf {\bibinfo {volume} {5}},\
  \bibinfo {pages} {4061} (\bibinfo {year} {2014})}\BibitemShut {NoStop}%
\bibitem [{\citenamefont {Pander}\ \emph {et~al.}(2018)\citenamefont {Pander},
  \citenamefont {Motyka}, \citenamefont {Zassowski}, \citenamefont
  {Etherington}, \citenamefont {Varsano}, \citenamefont {da~Silva},
  \citenamefont {Caldas}, \citenamefont {Data},\ and\ \citenamefont
  {Monkman}}]{monkman}%
  \BibitemOpen
  \bibfield  {author} {\bibinfo {author} {\bibfnamefont {P.}~\bibnamefont
  {Pander}}, \bibinfo {author} {\bibfnamefont {R.}~\bibnamefont {Motyka}},
  \bibinfo {author} {\bibfnamefont {P.}~\bibnamefont {Zassowski}}, \bibinfo
  {author} {\bibfnamefont {M.~K.}\ \bibnamefont {Etherington}}, \bibinfo
  {author} {\bibfnamefont {D.}~\bibnamefont {Varsano}}, \bibinfo {author}
  {\bibfnamefont {T.~J.}\ \bibnamefont {da~Silva}}, \bibinfo {author}
  {\bibfnamefont {M.~J.}\ \bibnamefont {Caldas}}, \bibinfo {author}
  {\bibfnamefont {P.}~\bibnamefont {Data}}, \ and\ \bibinfo {author}
  {\bibfnamefont {A.~P.}\ \bibnamefont {Monkman}},\ }\href {\doibase
  10.1021/acs.jpcc.8b07610} {\bibfield  {journal} {\bibinfo  {journal} {The
  Journal of Physical Chemistry C}\ }\textbf {\bibinfo {volume} {122}},\
  \bibinfo {pages} {23934} (\bibinfo {year} {2018})},\ \Eprint
  {http://arxiv.org/abs/https://doi.org/10.1021/acs.jpcc.8b07610}
  {https://doi.org/10.1021/acs.jpcc.8b07610} \BibitemShut {NoStop}%
\bibitem [{\citenamefont {Improta}\ \emph
  {et~al.}(2006{\natexlab{b}})\citenamefont {Improta}, \citenamefont {Barone},
  \citenamefont {Scalmani},\ and\ \citenamefont {Frisch}}]{improta2006}%
  \BibitemOpen
  \bibfield  {author} {\bibinfo {author} {\bibfnamefont {R.}~\bibnamefont
  {Improta}}, \bibinfo {author} {\bibfnamefont {V.}~\bibnamefont {Barone}},
  \bibinfo {author} {\bibfnamefont {G.}~\bibnamefont {Scalmani}}, \ and\
  \bibinfo {author} {\bibfnamefont {M.~J.}\ \bibnamefont {Frisch}},\
  }\href@noop {} {\bibfield  {journal} {\bibinfo  {journal} {J. Chem. Phys.}\
  }\textbf {\bibinfo {volume} {125}},\ \bibinfo {pages} {054103} (\bibinfo
  {year} {2006}{\natexlab{b}})}\BibitemShut {NoStop}%
\bibitem [{\citenamefont {Sissa}\ \emph
  {et~al.}(2012{\natexlab{a}})\citenamefont {Sissa}, \citenamefont {Calabrese},
  \citenamefont {Cavazzini}, \citenamefont {Grisanti}, \citenamefont
  {Terenziani}, \citenamefont {Quici},\ and\ \citenamefont
  {Painelli}}]{annine}%
  \BibitemOpen
  \bibfield  {author} {\bibinfo {author} {\bibfnamefont {C.}~\bibnamefont
  {Sissa}}, \bibinfo {author} {\bibfnamefont {V.}~\bibnamefont {Calabrese}},
  \bibinfo {author} {\bibfnamefont {M.}~\bibnamefont {Cavazzini}}, \bibinfo
  {author} {\bibfnamefont {L.}~\bibnamefont {Grisanti}}, \bibinfo {author}
  {\bibfnamefont {F.}~\bibnamefont {Terenziani}}, \bibinfo {author}
  {\bibfnamefont {S.}~\bibnamefont {Quici}}, \ and\ \bibinfo {author}
  {\bibfnamefont {A.}~\bibnamefont {Painelli}},\ }\href {\doibase
  10.1002/chem.201202154} {\bibfield  {journal} {\bibinfo  {journal} {Chemistry
  - A European Journal}\ }\textbf {\bibinfo {volume} {19}},\ \bibinfo {pages}
  {924} (\bibinfo {year} {2012}{\natexlab{a}})}\BibitemShut {NoStop}%
\bibitem [{\citenamefont {Anderson}(1996)}]{anderson}%
  \BibitemOpen
  \bibfield  {author} {\bibinfo {author} {\bibfnamefont {P.~W.}\ \bibnamefont
  {Anderson}},\ }\href@noop {} {\emph {\bibinfo {title} {Basic Notions of
  Condensed Matter Physics}}}\ (\bibinfo  {publisher} {CRC Press},\ \bibinfo
  {address} {Boca Raton, FL},\ \bibinfo {year} {1996})\BibitemShut {NoStop}%
\bibitem [{\citenamefont {Sissa}\ \emph {et~al.}(2014)\citenamefont {Sissa},
  \citenamefont {Delchiaro}, \citenamefont {Di~Maiolo}, \citenamefont
  {Terenziani},\ and\ \citenamefont {Painelli}}]{dimaiolo-jcp}%
  \BibitemOpen
  \bibfield  {author} {\bibinfo {author} {\bibfnamefont {C.}~\bibnamefont
  {Sissa}}, \bibinfo {author} {\bibfnamefont {F.}~\bibnamefont {Delchiaro}},
  \bibinfo {author} {\bibfnamefont {F.}~\bibnamefont {Di~Maiolo}}, \bibinfo
  {author} {\bibfnamefont {F.}~\bibnamefont {Terenziani}}, \ and\ \bibinfo
  {author} {\bibfnamefont {A.}~\bibnamefont {Painelli}},\ }\href {\doibase
  10.1063/1.4898710} {\bibfield  {journal} {\bibinfo  {journal} {The Journal of
  Chemical Physics}\ }\textbf {\bibinfo {volume} {141}},\ \bibinfo {pages}
  {164317} (\bibinfo {year} {2014})},\ \Eprint
  {http://arxiv.org/abs/https://doi.org/10.1063/1.4898710}
  {https://doi.org/10.1063/1.4898710} \BibitemShut {NoStop}%
\bibitem [{\citenamefont {Sissa}\ \emph
  {et~al.}(2012{\natexlab{b}})\citenamefont {Sissa}, \citenamefont {Jahani},
  \citenamefont {Soos},\ and\ \citenamefont {Painelli}}]{sissa:cyanine}%
  \BibitemOpen
  \bibfield  {author} {\bibinfo {author} {\bibfnamefont {C.}~\bibnamefont
  {Sissa}}, \bibinfo {author} {\bibfnamefont {P.~M.}\ \bibnamefont {Jahani}},
  \bibinfo {author} {\bibfnamefont {Z.~G.}\ \bibnamefont {Soos}}, \ and\
  \bibinfo {author} {\bibfnamefont {A.}~\bibnamefont {Painelli}},\ }\href
  {\doibase 10.1002/cphc.201200021} {\bibfield  {journal} {\bibinfo  {journal}
  {ChemPhysChem}\ }\textbf {\bibinfo {volume} {13}},\ \bibinfo {pages} {2795}
  (\bibinfo {year} {2012}{\natexlab{b}})},\ \Eprint
  {http://arxiv.org/abs/https://chemistry-europe.onlinelibrary.wiley.com/doi/pdf/10.1002/cphc.201200021}
  {https://chemistry-europe.onlinelibrary.wiley.com/doi/pdf/10.1002/cphc.201200021}
  \BibitemShut {NoStop}%
\bibitem [{\citenamefont {Katan}\ \emph
  {et~al.}(2005{\natexlab{a}})\citenamefont {Katan}, \citenamefont
  {Terenziani}, \citenamefont {Mongin}, \citenamefont {Werts}, \citenamefont
  {Porrès}, \citenamefont {Pons}, \citenamefont {Mertz}, \citenamefont
  {Tretiak},\ and\ \citenamefont {Blanchard-Desce}}]{terenziani2005a}%
  \BibitemOpen
  \bibfield  {author} {\bibinfo {author} {\bibfnamefont {C.}~\bibnamefont
  {Katan}}, \bibinfo {author} {\bibfnamefont {F.}~\bibnamefont {Terenziani}},
  \bibinfo {author} {\bibfnamefont {O.}~\bibnamefont {Mongin}}, \bibinfo
  {author} {\bibfnamefont {M.~H.~V.}\ \bibnamefont {Werts}}, \bibinfo {author}
  {\bibfnamefont {L.}~\bibnamefont {Porrès}}, \bibinfo {author} {\bibfnamefont
  {T.}~\bibnamefont {Pons}}, \bibinfo {author} {\bibfnamefont {J.}~\bibnamefont
  {Mertz}}, \bibinfo {author} {\bibfnamefont {S.}~\bibnamefont {Tretiak}}, \
  and\ \bibinfo {author} {\bibfnamefont {M.}~\bibnamefont {Blanchard-Desce}},\
  }\href {\doibase 10.1021/jp044193e} {\bibfield  {journal} {\bibinfo
  {journal} {The Journal of Physical Chemistry A}\ }\textbf {\bibinfo {volume}
  {109}},\ \bibinfo {pages} {3024} (\bibinfo {year} {2005}{\natexlab{a}})},\
  \bibinfo {note} {pMID: 16833626},\ \Eprint
  {http://arxiv.org/abs/https://doi.org/10.1021/jp044193e}
  {https://doi.org/10.1021/jp044193e} \BibitemShut {NoStop}%
\bibitem [{\citenamefont {Katan}\ \emph
  {et~al.}(2005{\natexlab{b}})\citenamefont {Katan}, \citenamefont
  {Terenziani}, \citenamefont {Droumaguet}, \citenamefont {Mongin},
  \citenamefont {Werts}, \citenamefont {Tretiak},\ and\ \citenamefont
  {Blanchard-Desce}}]{terenziani2005b}%
  \BibitemOpen
  \bibfield  {author} {\bibinfo {author} {\bibfnamefont {C.}~\bibnamefont
  {Katan}}, \bibinfo {author} {\bibfnamefont {F.}~\bibnamefont {Terenziani}},
  \bibinfo {author} {\bibfnamefont {C.~L.}\ \bibnamefont {Droumaguet}},
  \bibinfo {author} {\bibfnamefont {O.}~\bibnamefont {Mongin}}, \bibinfo
  {author} {\bibfnamefont {M.~H.~V.}\ \bibnamefont {Werts}}, \bibinfo {author}
  {\bibfnamefont {S.}~\bibnamefont {Tretiak}}, \ and\ \bibinfo {author}
  {\bibfnamefont {M.}~\bibnamefont {Blanchard-Desce}},\ }in\ \href {\doibase
  10.1117/12.618464} {\emph {\bibinfo {booktitle} {Linear and Nonlinear Optics
  of Organic Materials V}}},\ Vol.\ \bibinfo {volume} {5935},\ \bibinfo
  {editor} {edited by\ \bibinfo {editor} {\bibfnamefont {M.}~\bibnamefont
  {Eich}}},\ \bibinfo {organization} {International Society for Optics and
  Photonics}\ (\bibinfo  {publisher} {SPIE},\ \bibinfo {year} {2005})\ pp.\
  \bibinfo {pages} {13 -- 27}\BibitemShut {NoStop}%
\bibitem [{\citenamefont {Lukasiewicz}\ \emph {et~al.}(2020)\citenamefont
  {Lukasiewicz}, \citenamefont {Rammo}, \citenamefont {Stark}, \citenamefont
  {Krzeszewski}, \citenamefont {Jacquemin}, \citenamefont {Rebane},\ and\
  \citenamefont {Gryko}}]{jacquemin2020}%
  \BibitemOpen
  \bibfield  {author} {\bibinfo {author} {\bibfnamefont {L.~G.}\ \bibnamefont
  {Lukasiewicz}}, \bibinfo {author} {\bibfnamefont {M.}~\bibnamefont {Rammo}},
  \bibinfo {author} {\bibfnamefont {C.}~\bibnamefont {Stark}}, \bibinfo
  {author} {\bibfnamefont {M.}~\bibnamefont {Krzeszewski}}, \bibinfo {author}
  {\bibfnamefont {D.}~\bibnamefont {Jacquemin}}, \bibinfo {author}
  {\bibfnamefont {A.}~\bibnamefont {Rebane}}, \ and\ \bibinfo {author}
  {\bibfnamefont {D.~T.}\ \bibnamefont {Gryko}},\ }\href {\doibase
  10.1002/cptc.202000013} {\bibfield  {journal} {\bibinfo  {journal}
  {ChemPhotoChem}\ }\textbf {\bibinfo {volume} {4}},\ \bibinfo {pages} {508}
  (\bibinfo {year} {2020})},\ \Eprint
  {http://arxiv.org/abs/https://chemistry-europe.onlinelibrary.wiley.com/doi/pdf/10.1002/cptc.202000013}
  {https://chemistry-europe.onlinelibrary.wiley.com/doi/pdf/10.1002/cptc.202000013}
  \BibitemShut {NoStop}%
\bibitem [{\citenamefont {Kim}\ \emph {et~al.}(2020)\citenamefont {Kim},
  \citenamefont {Kim}, \citenamefont {Kang}, \citenamefont {Hong},
  \citenamefont {Würthner},\ and\ \citenamefont {Kim}}]{dongho2020}%
  \BibitemOpen
  \bibfield  {author} {\bibinfo {author} {\bibfnamefont {W.}~\bibnamefont
  {Kim}}, \bibinfo {author} {\bibfnamefont {T.}~\bibnamefont {Kim}}, \bibinfo
  {author} {\bibfnamefont {S.}~\bibnamefont {Kang}}, \bibinfo {author}
  {\bibfnamefont {Y.}~\bibnamefont {Hong}}, \bibinfo {author} {\bibfnamefont
  {F.}~\bibnamefont {Würthner}}, \ and\ \bibinfo {author} {\bibfnamefont
  {D.}~\bibnamefont {Kim}},\ }\href {\doibase 10.1002/anie.202005696}
  {\bibfield  {journal} {\bibinfo  {journal} {Angewandte Chemie International
  Edition}\ }\textbf {\bibinfo {volume} {59}},\ \bibinfo {pages} {8306}
  (\bibinfo {year} {2020})},\ \Eprint
  {http://arxiv.org/abs/https://onlinelibrary.wiley.com/doi/pdf/10.1002/anie.202005696}
  {https://onlinelibrary.wiley.com/doi/pdf/10.1002/anie.202005696} \BibitemShut
  {NoStop}%
\end{thebibliography}%

\end{document}